\documentclass[lettersize,journal]{IEEEtran}
\usepackage{amsmath,amsfonts}
\usepackage{algorithmic}
\usepackage{algorithm}
\usepackage{array}
\usepackage[caption=false,font=normalsize,labelfont=sf,textfont=sf]{subfig}
\usepackage{textcomp}
\usepackage{url}
\usepackage{verbatim}
\usepackage{graphicx}
\usepackage{cite}

\begin{document}

\title{Air Computing: A Survey on a New Generation Computation Paradigm in 6G Wireless Networks}

\author{Baris Yamansavascilar, Atay Ozgovde, and Cem Ersoy}

\maketitle

\begin{abstract}
There is an ever-growing race between what novel applications demand from the infrastructure and what the continuous technological breakthroughs bring in. Especially after the proliferation of smart devices and diverse IoT requirements, we observe the dominance of cutting-edge applications with ever-increased user expectations in terms of mobility, pervasiveness, and real-time response. Over the years, to meet the requirements of those applications, cloud computing provides the necessary capacity for computation, while edge computing ensures low latency. However, these two essential solutions would be insufficient for the next-generation applications since computational and communicational bottlenecks are inevitable due to the highly dynamic load. Therefore, a 3D networking structure using different air layers including Low Altitude Platforms, High Altitude Platforms, and Low Earth Orbits in a harmonized manner for both urban and rural areas should be applied to satisfy the requirements of the dynamic environment. In this perspective, we put forward a novel, next-generation paradigm called Air Computing that presents a dynamic, responsive, and high-resolution computation and communication environment for all spectrum of applications using the 6G Wireless Networks as the fundamental communication system. In this survey, we define the components of air computing, investigate its architecture in detail, and discuss its essential use cases and the advantages it brings for next-generation application scenarios. We provide a detailed and technical overview of the benefits and challenges of air computing as a novel paradigm and spot the important future research directions.

\end{abstract}


\section{Introduction}








\IEEEPARstart{I}{n} order to meet the stringent demands of fully connected, intelligent, and computation-intensive applications with low latency support, sixth generation (6G) wireless networks provide vertical networking solutions \cite{mao2021optimizing}. Especially, considering the number of Internet of Things (IoT) connections which are estimated as 14.7 billion in 2023, vertical networking would be crucial for the seamless coverage and dense connection capabilities \cite{saad2019vision, giordani2020toward}. Moreover, since the connection of devices that have separate requirements must be processed heterogeneously to ensure Quality of Experience (QoE), reliable computation of different task types is critical in the next-generation communication systems \cite{khan2017reliable}. 

Traditional terrestrial communications including the fifth generation wireless networks (5G) provide the infrastructure for applications of end-users. Those applications demand diverse requirements which are hard to satisfy with well-known practices \cite{li2021understanding}. Therefore, deployment of Unmanned Aerial Vehicles (UAVs) as low altitude platforms (LAP), airplanes as high altitude platforms (HAP), and low earth orbit (LEO) satellites are legitimate candidates for future networks in order to satisfy the requirements of different applications since they can provide low latency, high computation capability, reliability, and availability. These properties are specifically important for the processing of the corresponding tasks of those applications in terms of content caching, resource allocation, task offloading, and extreme mobility.

Although edge computing provides promising results in urban settings in the short-run, reaching genuine ubiquitous execution of real-time, computationally intensive novel applications will require further approaches. Air computing in this respect harmonizes traditional terrestrial edge computing with a wide range of air technologies to obtain a robust, high-capacity computational infrastructure that embraces urban, suburban, and rural scenarios.


\subsection{Air Computing as a New Computational Paradigm}

In the literature, the organization and orchestration of the 3D networking structure is called under different names such as aerial communication, Space-air-ground Integrated Network (SAGIN), and airborne networks \cite{guo2021service, liu2018space, baltaci2021survey}. Especially, the aerial term is widely used to point to the utilization of air components in 3D networking. Since air layers and the corresponding air components would be an essential part of the next-generation communication systems rather than an auxiliary, we call this next-generation computation paradigm as air computing.

\begin{figure*}[!t]
\centering
\includegraphics[scale=0.1]{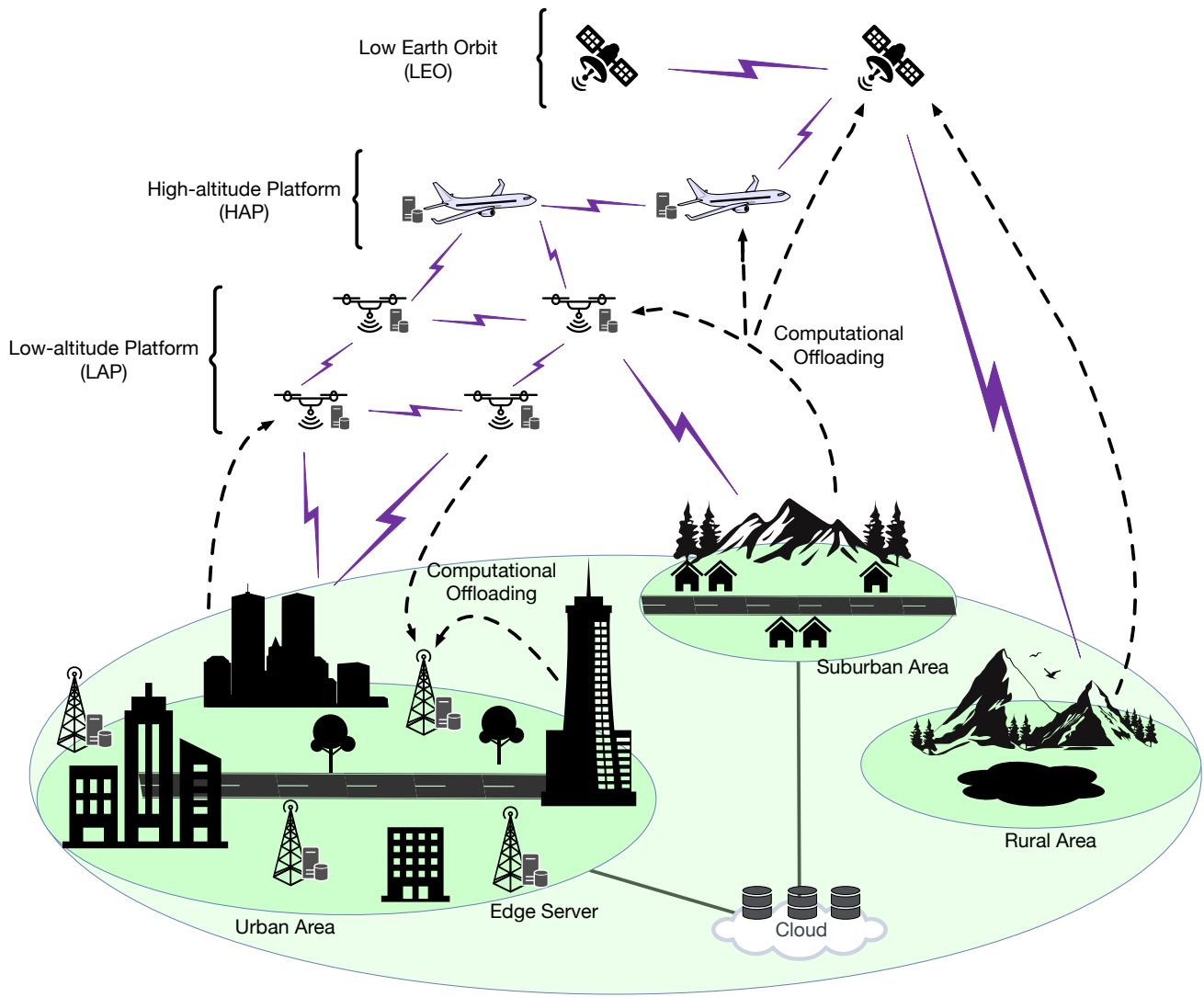}
\caption{Air Computing Architecture.}
\label{AirComputingGeneral}
\end{figure*}

\subsection{Motivation}

Even though air computing can be theoretically applicable with current 5G wireless networks, it is not suitable to carry out completely since 5G infrastructure does not offer ultra-low latency (0.1 ms), ultra-high data rate (1 Tbps), and mobility as 1000 km/hr, which are essential for mission critical applications and many future applications \cite{huynh2021envisioning}. Moreover, seamless coverage, which is crucial for air computing considering the communication between terrestrial systems and air layers, cannot be sufficiently provided by 5G due to its network features \cite{gupta20216g}. Therefore, 6G communication systems are the genuine candidate for air computing as they ensure those essential features \cite{masaracchia2021uav}. Moreover, since edge intelligence will be a core part of 6G communication architecture, artificial intelligence (AI) would be used prevalently as the solution for computing processes of applications \cite{jagannath2021redefining}.

As the capabilities of 6G wireless networks allow to satisfy the strict demands of modern applications, a wise deployment of terrestrial servers and air components for the computational needs will change the traditional methods such as edge and cloud computing. Therefore, in this study, we investigate the computing opportunities that would be the result of the intelligent communication between terrestrial servers and air vehicles which we call air computing. These opportunities manifest themselves as an enhancement of QoS and QoE considering both communication requirements and end-user needs  \cite{sisinni2018industrial, siddiqi20195g}. Moreover, as shown in Figure \ref{AirComputingGeneral}, we believe that the coordination between devices and different communication mediums through the air would open new research challenges that will shape the future of the Internet.

We foresee that air computing will be the next generation computation paradigm as a result of the evolution of multi-access edge computing (MEC) \cite{wang2018power}. Since the computation paradigm in MEC is typically restricted with the 2D terrestrial networks in which the resources allocated for the application tasks in either an edge server or cloud server, system bottlenecks can limit meeting the diverse requirements of heterogeneous applications. Since air computing is comprised of different air communication technologies, resource allocation alternatives are considerably increased when compared with the 2D settings. Vertical networking structure shown in 
Figure \ref{AirComputingGeneral} depicts how different applications would use different resources to ensure their QoS requirements.


\subsection{Research Scope and Contributions}
In this survey, we focus on the computational requirements of the next generation heterogeneous applications and corresponding solutions formulated as the air computing in which terrestrial servers, LAP, HAP, and LEO layers are coordinated intelligently. To put relevant technologies in perspective, we first investigate the differences between edge computing and air computing in terms of the network architecture, challenges, and use cases. Next, we evaluate the advantages of air computing regarding latency, computation capability, storage, mobility, coverage, and reliability.

Since air computing includes both terrestrial servers and air components, we also examine studies on edge computing, UAVs, and other air components in order to show the benefits of air computing more concretely. Furthermore, we detail the possible scenarios in which air computing would be the only valid alternative.

It is important to note that the technical aspects of the aerial radio access network (ARAN) technologies which can be used in air computing are out of the scope of this survey. We only focus on the computation part of the air computing regarding the aforementioned architectural advantages and possible use cases. Regarding our air computing definition, there are three recent survey papers including \cite{cao2018airborne, baltaci2021survey, dao2021survey} that investigated the aerial radio access networks (ARANs). In \cite{cao2018airborne}, Cao et al. investigated the  mechanisms and protocols for airborne communication networks. They detailed LAP-based communication networks, and HAP-based communication networks regarding their channel models, protocols, and spectrum efficiency. In \cite{baltaci2021survey}, Baltaci et al. focused on the connectivity requirements and use cases of aerial vehicles considering the challenges of employing wireless communication standards. They introduce the term Future Aerial Communications (FACOM) for aerial connectivity and its use cases. They also examined Radio Frequency (RF) wireless technologies to apply in FACOM. In \cite{dao2021survey}, authors focused on the future network design, system model analysis, and enabling technologies in terms of 6G access infrastructure, transmission propagation, communication latency, and energy consumption. They defined the radio access model as ARAN. 


	
The main contributions of this survey are as follows. 

\begin{itemize}

\item We introduce air computing which is the next-generation computation paradigm for 6G wireless networks. We define its components including terrestrial, LAP, HAP, and LEO layers and investigate them thoroughly.


\item We analyze recent studies that focus on edge computing, UAVs, and other air components in the literature and compare them with air computing in order to show its concrete advantages and possible solutions that cannot be offered by traditional networking paradigms. 

\item We detail the scenarios for air computing that improve the QoS and QoE for end-users. Especially, we focus on mission-critical applications that may require 1 ms latency which may not be met by traditional computing schemes regarding Metropolitan Area Network (MAN) and Wide Area Network (WAN). 

\item We investigate the open research problems and challenges for air computing considering the architecture design, request management, utilization of artificial intelligence, a required protocol, energy issues, air regulations, and movement mechanisms. We believe that we point out important spots in the literature so that readers would expand their studies through one of those areas.

\end{itemize}

The rest of the paper is organized as follows. In Section II, we introduce air computing considering its advantages, and its differences with edge computing. We show the use cases of air computing in different scenarios in Section III. In Section IV, we investigate edge, LAPs, HAPs, and LEOs which are the main components of air computing. We examine corresponding studies and show the reader that unresolved issues in those papers would be solved by air computing. In Section V, we provide the challenges, opportunities, and future research directions. Finally, we conclude our paper in Section VI.

\section{Air Computing}

Air computing is a next-generation computational paradigm in which
ubiquitous applications with radical networking and computational requirements are satisfied with the help of a family of novel communication opportunities. It provides a highly dynamic, scalable and responsive computational infrastructure in which terrestrial servers are harmonized with various air layers including LAP, HAP, and LEO as shown in Figure \ref{AirComputingGeneral}. Moreover, air computing augments traditional 2D edge computing with a wide spectrum of different computational servers in the air considering a highly dynamic context.

Currently in the urban area, users can enjoy the underlying terrestrial resources to experience seamless connection based on the available infrastructure. With the help of edge computing, mobile devices can reach one of the nearest servers for the execution of their delegated tasks via offloading mechanisms. Despite these advanced approaches, ever-growing application requirements and increasing user mobility patterns push the limits of the fixed infrastructure which led to UAVs being deployed for dynamic capacity enhancement \cite{kurt2021vision, cheng2018air}. Adding a vertical dimension to the network greatly enhances the possibilities in terms of the interaction of the users with computational resources. Accordingly, one of the main features of next-generation systems is expected to be their dynamically provisioned 3-dimensional (3D) structure which leads to many opportunities in terms of QoS and user experience \cite{li2018uav}. In a typical edge computing scenario, computational offloading would end either in an edge server or in a cloud server in the 2D terrestrial networks. This setting requires solving an NP-hard optimization problem \cite{chen2015efficient} due to the long term selection strategy. Cloud servers, although providing seemingly infinite computational capacity, due to latency may be prohibitive or cause low QoE. In that respect, by adding a new dimension using UAVs and other HAP entities, the overall capacity is considerably enhanced and access becomes agile which leads to the server selection process to be more versatile. Since different application types would have instant access to the dynamically arranged computational array of resources in the air, as well as terrestrial ones, this architecture will provide a dramatically increased QoE offerings. Moreover, air computing will also address users or autonomous entities in the air. A user in an air vehicle can perform computational offloading to other air units and/or terrestrial servers, which manifests itself as a capacity enhancement. As shown in Figure  \ref{AirComputingGeneral}, the offloading can be directly from the air vehicle or indirectly via routing over a LAP or a HAP before it is sent to the corresponding server. 

A suburban area consists of residential homes and has less population density than the urban area. Hence, the communication infrastructure is not as pervasive as in the urban area which results in fewer resources for the computational needs of the applications. Even though the cloud servers can be reachable via existing infrastructure, the latency would be high for many applications which require low latency for the QoS. Therefore, using LAP and HAP layers as the vertical networking for the suburban areas will increase QoE. The first effect would be on the capacity of the area as new resources can be available for the applications of the users. The second influence would be on the latency since time-critical applications can use the computational sources of LAP vehicles for their corresponding tasks without using the cloud resources that cause high delay. Third, the coverage in the area can be enhanced by placing the UAVs into the corresponding places and the connection would not be interrupted \cite{kurt2021vision, el2021uav}.

In the rural area, we assume that there is an environment in which people perform different activities such as sailing, kayaking, climbing, trekking, and camping. The essential fact is that the communication infrastructure may not exist or exist with extremely limited capacity. Moreover, accessing the cloud server is not always possible. In these circumstances, air computing can be used to meet essential QoS requirements since users would utilize fundamental resources for their applications. Even though LAP and HAP platforms are the backbone of the air computing, LEOs are generally used in rural areas since the replacement and energy refilling could be important issues for air vehicles such as UAVs. By deploying LEOs as the computational server or as the relay node that offloads the corresponding tasks to the suitable server, the end-user can enjoy the benefits of the applications also in rural areas.

\subsection{Differences Between Air and Edge Computing}
\begin{figure}[t]
\centering
\includegraphics[scale=0.35]{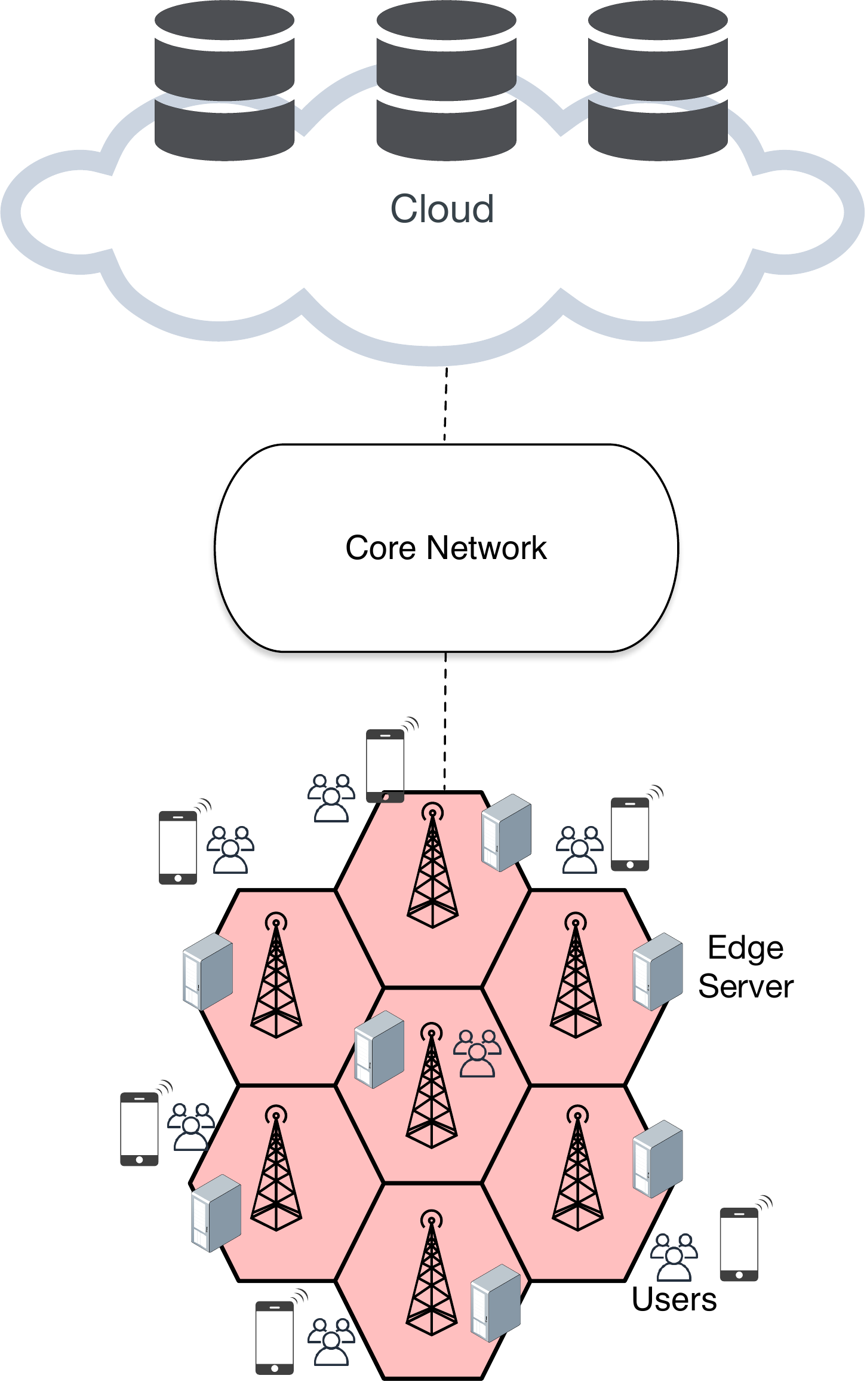}
\caption{Edge Computing.}
\label{EdgeCompGeneral}
\end{figure}

Since several application types such as image rendering, video editing, and simulation require intensive computation, cloud computing is proposed as the possible solution regarding CPU and battery constraints of end-devices \cite{senyo2018cloud}. However, with the proliferation of versatile devices and corresponding applications known as the Internet of Things (IoT), cloud computing could not meet the low latency requirements \cite{laroui2021edge}. As a result, several computation architectures including Cloudlet \cite{satyanarayanan2009case}, Multi-access Edge Computing (MEC) \cite{chen2015efficient}, and Fog Computing \cite{computing2016fog} emerged considering the replacement of the servers, which are not as powerful as the cloud servers, into the local area network (LAN). In the literature, these architectures are named under the umbrella term edge computing \cite{baktir2017can} which is also used in this study. The general architecture of edge computing is shown in Figure \ref{EdgeCompGeneral}. 

The main idea behind the edge computing is that processing the computation-intensive tasks, which cannot be processed in the end-device due to CPU and battery limitation, in the suitable server in the LAN. Moreover, the concept can be enhanced to metropolitan area network (MAN) due to scarce capacity \cite{laroui2021edge}. In that case, the most suitable server in the MAN is selected for the corresponding application. Furthermore, cloud servers can also be used with edge computing for latency insensitive applications in order to serve more users. However, selection of the most suitable edge or cloud server for the long-term optimization would be NP-hard \cite{chen2015efficient}. In general, edge computing is used for many application domains such as agriculture, healthcare, smart home, robotics, data processing, video analytics, and virtual reality \cite{mach2017mobile, mao2017survey}.

5G systems offers low-latency under 5 ms but it is not sufficient to provide ultra-low latency which is 0.1 ms, high mobility that is 1000 km/hr, and high connection density which is $10^7$ devices/sq km \cite{nguyen20216g}. As a result, although edge computing is a crucial networking paradigm in terms of QoS and QoE for diverse application needs, the requirement of recent mission-critical applications that require ultra low-latency may not be met by the existing edge infrastructure that uses 5G wireless networks. Moreover, the capacity of the networks considering 2D terrestrial resources is limited to serve very dense mobile and IoT devices. To solve these issues, air computing provides a third layer which is the air including LAP, HAP, and LEO layer to enhance the 2D computational paradigm into 3D by using the technological advantages of 6G communication systems. As 6G networks provide end-to-end latency under 1 ms, air latency between 10-100 $\mu$s, and reliability of $10^{-9}$, it meets future application requirements as an infrastructure regarding computation paradigms such as air computing \cite{masaracchia2021uav}. The 3D structure on the other hand is also considered as vertical networking and it provides important solutions that cannot be given by traditional edge computing. Thus, air computing can also be seen as the evolution of  edge computing in 6G communication systems.

\begin{figure}[t]
\captionsetup[subfigure]{font=scriptsize,labelfont=scriptsize}
\centering
\subfloat[][\label{edgeHorizontal}Horizontal and one-way direction in edge computing]{
\includegraphics[scale=0.075]{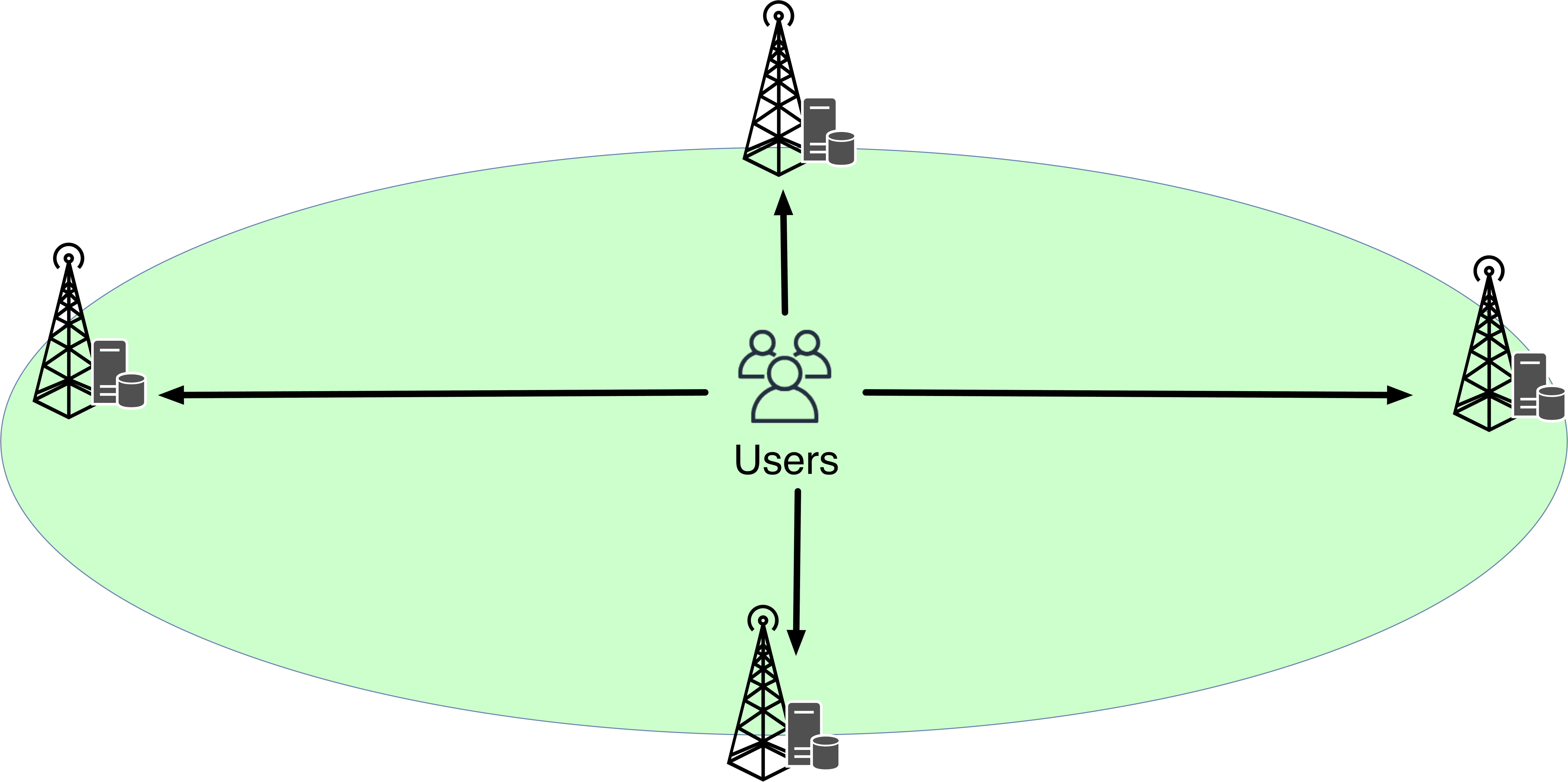}
}\\
\subfloat[][\label{airVertical} Horizontal, vertical and two-way direction in air computing ]{
\includegraphics[scale=0.075]{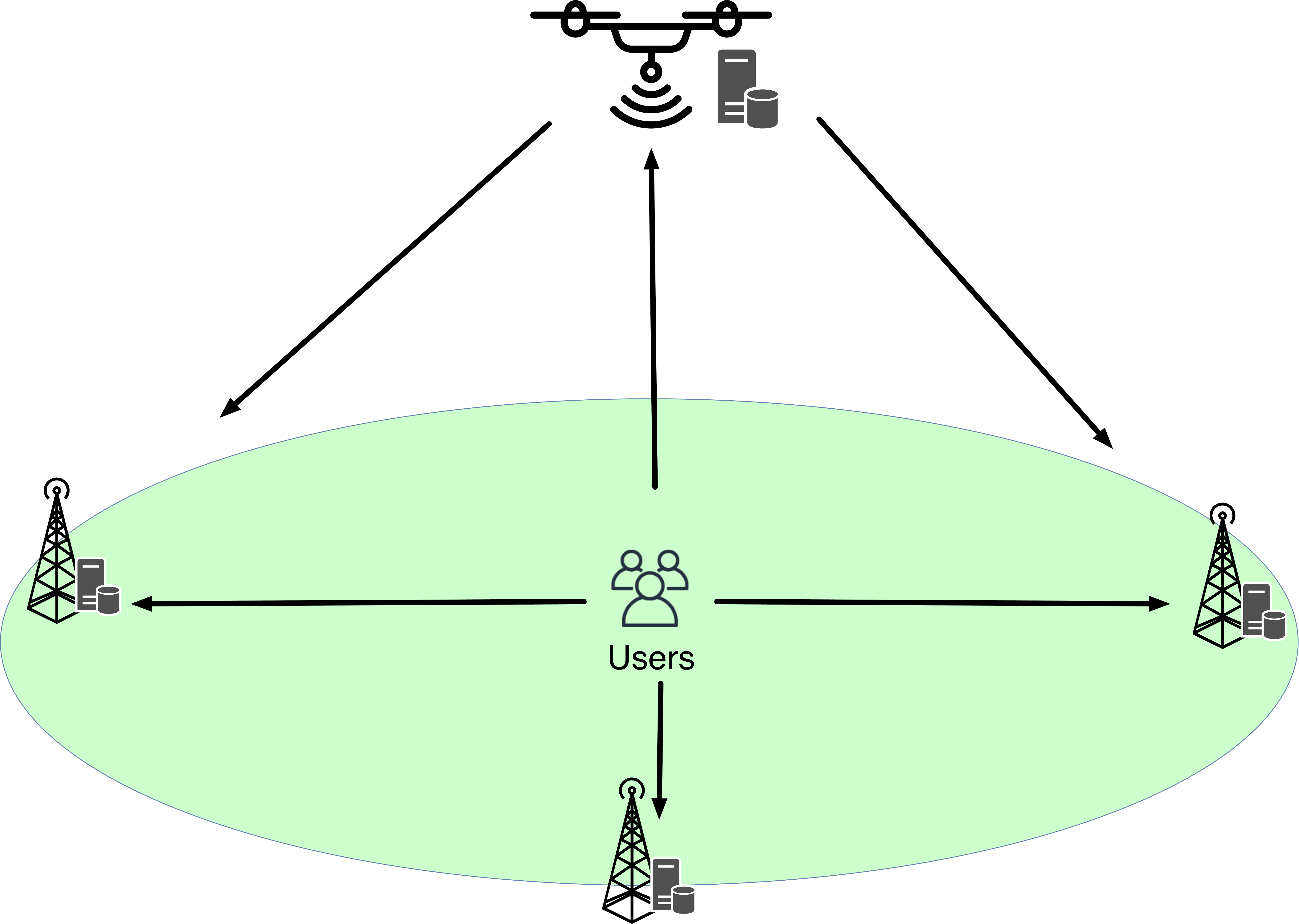}
}\\
\caption{Direction of the computational task offloading in air and edge computing.} \label{EdgeVsAir}
\end{figure}

As shown in Figure \ref{EdgeVsAir}, one of the most important differences between edge and air computing is the direction of the computational task offloading. In edge computing, the direction is horizontal as computational resources are inside the terrestrial 2D area. Moreover, these directions are always one way that is from the user device to the corresponding server. An offloading of a task or a request comes from the user application to the server and then the server process the task in order to give the suitable service. On the other hand, in air computing, the direction of the computational offloading is both horizontal and vertical. An application would benefit from terrestrial resources and air platforms at the same time. Furthermore, the vertical case of the computational offloading may be in both directions. A typical user in an urban area can use the resources in the air and, conversely, an application in an air vehicle can also benefit from the terrestrial servers. The important differences between edge and air computing are summarized in Table \ref{EdgeAndAirDiff}.

\subsection{Advantages of Air Computing}

Air computing has many advantages that can be categorized as {\itshape latency, data rate, computational capability, coverage,} and \textit{mobility}. In this subsection, we explain those advantages in detail.

\subsubsection{Task Offloading}
The main advantage of air computing regarding task offloading is the vertical network opportunities. Unlike traditional edge computing which is based on terrestrial resources, air computing employs a wide variety of computational technologies in air layers each with a different degree of geographical and mobility capabilities. Moreover, air components not only add a new physical dimension to the overall infrastructure but also their ability to be dynamically arranged creates a vision where the system can swiftly adapt itself to the ever-changing conditions of the users, applications, and the network itself. This allows air computing to adequately respond to the full spectrum of application profiles including ones with stringent latency and computational requirements.


We can detail the advantages of air computing for task offloading in three different scenarios. In the first scenario, we can assume that a user device in a terrestrial place decides to offload an atomic task of an application which cannot be processed in the device itself. Hence, the task can be offloaded to either terrestrial resources or air components. If terrestrial servers are selected for the offloading, the corresponding procedure would be similar to the edge computing process in which the task is processed on the server and then the results are transmitted to the user. However, in contrast to edge computing, if edge servers in the terrestrial area cannot serve due to their limited capacity, the tasks can be offloaded to air components rather than the cloud servers. This is a crucial advantage of air computing since blockage-free air routes and dynamically provisioned servers would provide lower latency.

\begin{figure}[t]
\centering
\includegraphics[scale=0.15]{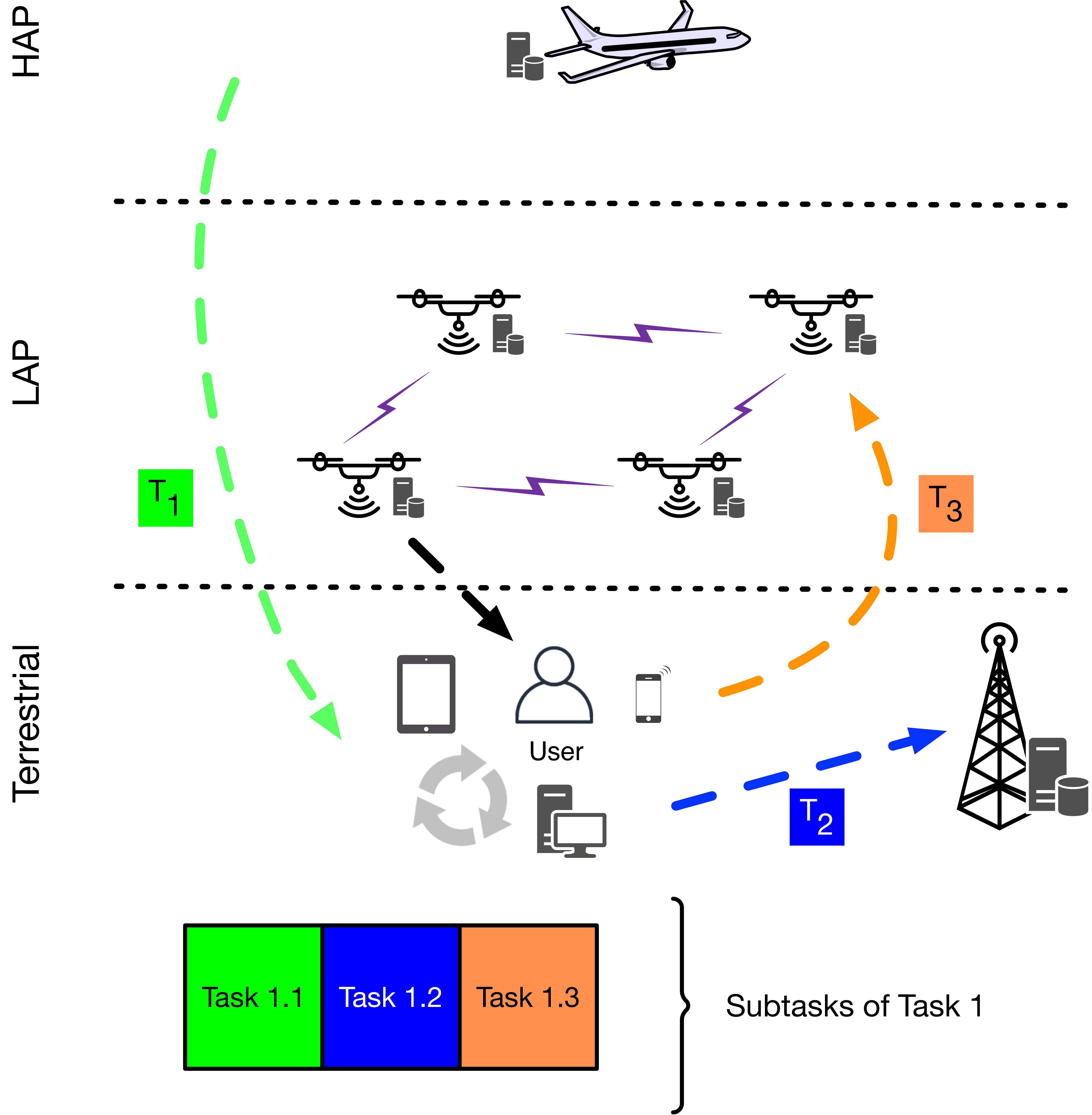}
\caption{Subtasks of a single task can be processed by different components of an air computing environment.}
\label{Subtasks}
\end{figure}

In the second scenario, we can assume that tasks are non-atomic and different parts of the main task can be processed in different resources in an air computing environment as shown in Figure \ref{Subtasks}. Since coverage is not a problem in vertical networking as in the case of traditional 2D networking, the corresponding partial tasks may be sent to the terrestrial resources in other regional domains if capacity problems occur in the air. As the communication would be in very high bit rates and the air resources would be in the vicinity, the latency may not be an important issue in this situation. Thus, partial offloading can be carried out more efficiently in air computing. On the other hand, if only the terrestrial resources are used for this purpose, this scenario can cause capacity problems regarding the processing if the number of users and their corresponding tasks are high with respect to the resources. Moreover, sending partial tasks to different terrestrial resources may cause congestion in particular parts of the network regarding the link capacity and user density. 


The third scenario for task offloading in air computing is that the tasks can be offloaded from the air to the ground. This is crucial since without air computing, users in the air must use the device capabilities, resources of the air vehicle, or relay capabilities of GEO satellites which results in low QoE. Now, through air computing, the tasks can be offloaded to terrestrial sources with low latency and applications can enjoy the benefits of the corresponding resources. Moreover, since air components can cover a large area, different tasks may be offloaded to different terrestrial areas. 

\begin{table}[t]
\caption{Important Differences between edge and air computing}
\label{EdgeAndAirDiff}
\centering

\begin{tabular}{ | p{3cm} | l | p{2cm} | p{2cm} |}
\hline
\textbf{Feature} & \textbf{Edge Computing} & \textbf{Air Computing}\\
\hline
Network Architecture & 2D terrestrial & 3D vertical\\
Typical Latency & $<$ 10 ms & $<$ 1 ms\\
Mobility & $<$ 500 km/hr & $<$ 1000 km/hr\\
Coverage & Cell-based & Cell-less\\
Offloading Direction & Horizontal one-way & Vertical two-way\\
Bandwidth & Static & Dynamic\\
\hline
\end{tabular}

\end{table}

\subsubsection{Content Caching}

Content caching is one of the important practices in order to access the requested pages, tools, and applications with low latency. Therefore, content caching optimization contains three objectives including QoS guarantee, content popularity, and utility maximization \cite{ouyang2018follow, zhang2018hierarchical}. To this end,  hit ratio is used as the primary metric to indicate the quality of the content caching optimization. Especially, when the storage capacity of the corresponding servers is insufficient, the quality of the optimization would be more important.

By using vertical networking through air computing, the capacity is enhanced regarding two possible methods. First, storage of the air vehicles can be used for this purpose. As a result, the capacity can be improved and more content can be reachable by users. In the second method, air components would be used as a relay for the request and the corresponding content can be reached through nearby terrestrial servers. This method can provide lower latency than using WAN. 

Even though air computing has important features for content caching in the urban area, its main advantage manifests itself in suburban and rural areas as the communication infrastructure is less developed. Considering the fact that even cloud resources cannot be reachable in rural areas, the importance of air computing would be better recognized. However, since using UAVs would be less efficient as they need corresponding battery charging stations which cannot be found in the rural area, HAPs and LEOs are more suitable to use. As both air platforms can provide content caching, the QoE of users would be enhanced.

\subsubsection{Latency}

\begin{figure}[t]
\centering
\includegraphics[scale=0.033]{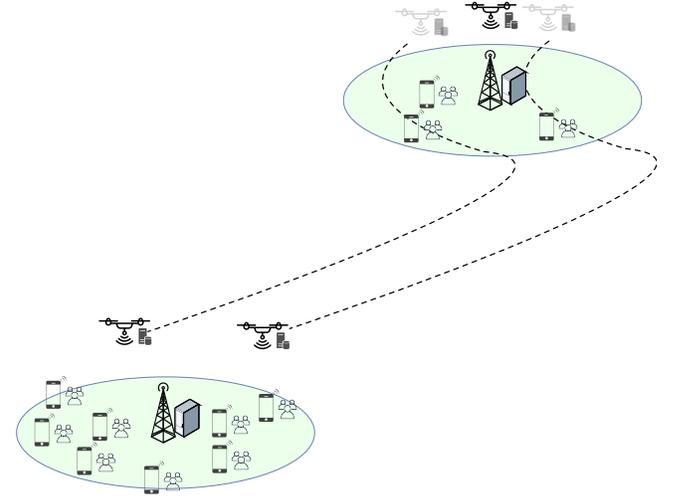}
\caption{Air components can be replaced in the environment dynamically based on the changing user demands in particular locations.}
\label{DynamicUAV}
\end{figure}

Regarding the QoS, one of the most important metrics for a computation paradigm is latency. Since users would like to obtain the corresponding content or result as quickly as possible, providing low latency is critical. Air computing makes use of vertical networking opportunities and 6G communication infrastructure to obtain latency figures under 1ms for specific scenarios. This allows air computing to support mission-critical as well as tactile application profiles such as remote health, mobile augmented reality, natural disaster emergency intervention, and telesurgery. In traditional terrestrial networking paradigms such as edge computing, the range of the servers is crucial for the latency even though edge servers are located at the LAN or MAN. On the other hand, as air computing allows for the dynamic placement and provisioning of resources in places needed, typically latency would be independent of the geographical location of the users as shown in Figure \ref{DynamicUAV}. This advantage also provides important stability in terms of QoE.

\subsubsection{Data Rate}

Especially for streaming applications, model training, and content sharing, the data rate is crucial. As 6G networks will provide up to 1 Tbps data rate, the upload and download rate of corresponding applications would be higher \cite{inomata2021terahertz}. Hence, the throughput and QoE of traditional and mission-critical applications would be increased \cite{tariq2020speculative}. Moreover, it allows air computing to handle the computational requirements of the next-generation applications such as holographic telepresence which requires high bandwidth. 

\subsubsection{Computational Capability}
Even though the computational power of the air vehicles is not as powerful as the terrestrial servers, the tasks of applications can be offloaded to multiple sources to enhance the throughput. Moreover, by using both terrestrial servers and air components, a single task may be partitioned to be processed. Since the data rate is higher and latency is ultra-low in air computing regarding edge computing, using several sources in different layers for a single task would increase QoE. 

\subsubsection{Coverage}

\begin{figure}[t]
\centering
\includegraphics[scale=0.07]{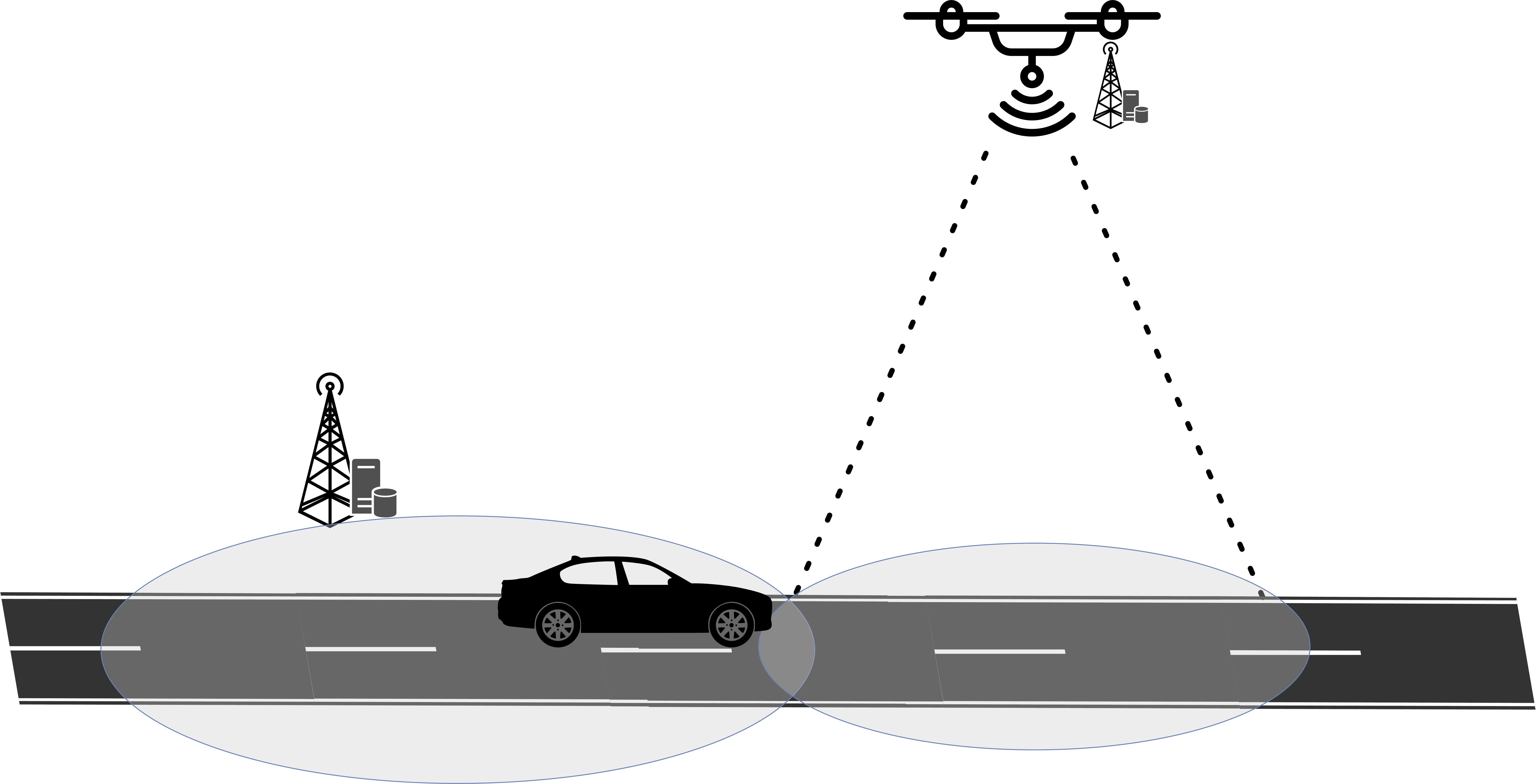}
\caption{Cell-less structure through air components would provide seamless connection for end-users.} 
\label{CellessStructure}
\end{figure}

Since air vehicles and their corresponding components are used in air computing, the end devices will not depend on the cell infrastructure in which there is a limited capacity regarding the number of users. As shown in Figure \ref{CellessStructure}, this cell-less structure will provide pervasive connectivity which is crucial for the heterogeneity of future applications.

\subsubsection{Mobility}

As a result of seamless coverage and pervasive connectivity, air computing provides high mobility which is over 1000 km/hr. Moreover, since air components communicate with each other, handover for the processed tasks of users that are in the air or terrestrial vehicle is carried out more smoothly. Furthermore, mobility in air computing can be considered for the users in the air and ground. Accordingly, the processed tasks can also be sent to the terrestrial servers from the air or vice versa.



\section{Air Computing Use Cases}

The actions that can be taken in air computing are similar to edge computing as they include computation/task offloading, resource allocation, and resource provisioning described in Section III. However, since air components provide important flexibility regarding vertical networking, the use cases in air computing are more versatile than edge computing. To this end, we elaborate on the potential use cases of air computing in this section. 

\subsection{Natural Disasters}


Natural disasters such as earthquakes, hurricanes, tsunami, and floods wreak havoc on settlements and residential areas. They cause the loss of human lives and the destruction of important resources that people need. Considering communication perspective, there would be two important consequences. First, the communication facilities would be destructed by the natural disaster and as a result people can be deprived of important resources that cause isolation in the disaster site. This is crucial for those people affected by the disaster because they cannot obtain the required aid which must be given to the heavy injury cases and also to humans that expect to be rescued. Without communication resources, the outcome of the disaster would be much worse. Secondly, even in the cases where communication facilities are not seriously damaged, bursty traffic caused by people that want to make an emergency call and to reach their friends, and relatives brings about congestion in the network. Note that this congestion can be also caused by the people outside the disaster site since they also would like to reach the people that are affected by the disaster. Similar to the first case, as a result, people can be isolated in the disaster site so that the outcome of the disaster would be heavier.

Considering both cases, the essential requirement for providing the communication and computation in a disaster site is to enhance the capacity dynamically. This can be carried out by using air components of air computing including UAVs, HAP vehicles, and LEOs. Note that the utilization of those components depends on the disaster type. For example, if the disaster is a hurricane, UAVs and HAP vehicles cannot be used during the disaster. Similarly, if the disaster is an earthquake or a flood, using UAVs would be more effective considering the latency which is a crucial metric for these scenarios.


\subsection{Mission Critical Applications}

Even though networking paradigms propose several solutions to mission-critical applications that require ultra-low latency, they may not always prove successful because of the underlying technology and incapability of terrestrial resources. Thanks to 6G wireless networking capabilities and air computing, the requirements of mission critical applications for a desired QoE can be met. In this section, we list some of those applications to show their benefits.

\subsubsection{Well-being Monitoring} 

As elderly people that suffer due to important diseases such as parkinson, epilepsy and alzheimer must be regularly monitored in the case of fall, crisis, and lost, high latency would be fatal for them especially in rural areas. Therefore, well-being monitoring must be provided such that the latency should not be destructive for the patients. As terrestrial resources cannot ensure ultra-low latency which is 1 ms, air computing can be used for this purpose. For the urban areas, air computing may be utilized as the complementary resource regarding capacity since 6G wireless networks can handle most of the requests. On the other hand, for suburban and rural areas, air computing would be the primary resource for well-being monitoring. By using the coverage, latency, and data rate advantages of air computing, the quality of life (QoL) of those patients can be enhanced.

\subsubsection{Remote Health}

The issues in remote health are similar to those in well-being monitoring, however they are more critical as medical operations are carried out in this case. The scope of remote health includes the actions of doctors and monitoring the results of those actions on the patients in operations. Moreover, critical life-related metrics such as heart beating, and adrenalin level must be constantly monitored in the operation. To this end, air computing provides important opportunities in this area through its paradigm and corresponding components. For example, if the patient or doctor cannot move from one place to another due to several reasons, the operation can be carried out from a remote area where the doctor leads it.


\subsubsection{Real-time Video}

Since real-time video requires a constant bit rate through the lifetime of the video, ensuring the desired QoE is more difficult than the video on demand systems in which the delay can be compensated using dynamically changing buffers. This issue is experienced differently by two main use-cases: (1) real-time conferences/calls such as Zoom and Skype, and (2) watching sports activities.

The fundamental problem in the real-time video for sports events through the Internet is that viewers obtain the content with higher delay regarding terrestrial broadcast. As a result, QoE reduces significantly as viewers may hear the sound of terrestrial broadcast viewers when an important incident has occurred in the event including a football or basketball competition. On the other hand, in conferences and calls, the video can stall or the voice cannot be synchronized with the video due to jitter or congestion in the network.

\begin{figure}[t]
\centering
\includegraphics[scale=0.15]{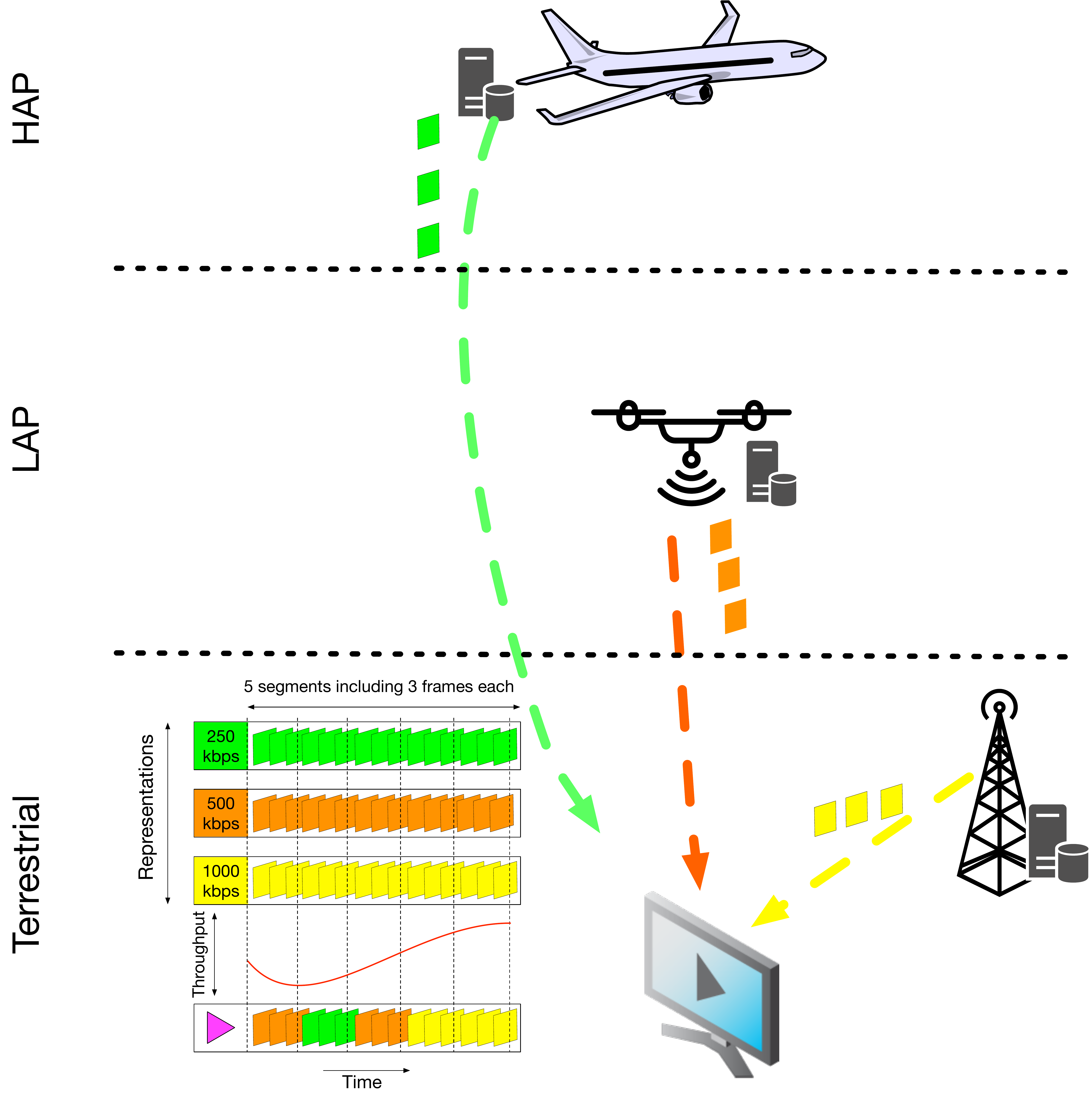}
\caption{Different representations of the video segments can be conveyed by different components of air computing based on the network conditions.}
\label{VideoAirCmp}
\end{figure}

Air computing can provide two possible solutions to those problems. The first solution is related to its high capacity in which Terahertz communication would reduce the latency significantly and therefore ensure to control the jitter. The second solution as shown in Figure \ref{VideoAirCmp} proposes a novel architecture for video streaming using air components. In this solution, the video segments are routed via different components including terrestrial servers, UAVs, HAP vehicles, and LEOs based on the current requirements of the network and video. Note that this approach may bring about its own challenges considering scheduling and management of segments and user profiles. However, by using the underlying technology and novel utilization of the air components, the problems above can be solved.

\subsubsection{Mobile Augmented Reality}
Ultra-low latency and extremely high data rates are crucial factors for augmented reality (AR) in which the real and virtual objects are combined \cite{chatzopoulos2017mobile}. Therefore, AR has been used in many application areas including entertainment, healthcare, and education \cite{manuri2016survey}. Moreover, since they perform in real-time, augmented reality applications are delay intolerant in order to provide satisfactory QoE for end-users. Especially, after the widespread usage of smartphones with their expanded capabilities, augmented reality applications are widely deployed. To this end, AR is recently named as mobile augmented reality (MAR).

The most significant difference between AR and MAR is their processing methods. While traditional AR applications perform in-device processing, MAR uses offloading to carry out corresponding computations \cite{siriwardhana2021survey}. The main reason for offloading in MAR is the limited capacity of mobile devices regarding battery and CPU. Therefore, edge and cloud computing have been broadly utilized by MAR applications in recent years. However, the required latency based on the application areas including healthcare, smart city, and Industry 4.0 has changed as the communication technologies are evolved such as 5G and 6G. As a result, the ultra-low latency requirement of those applications considering the mobility in MAN may not be met by edge solutions.

Air computing can solve the issues related to MAR including scalability, latency, and expected data capacity indicated in \cite{siriwardhana2021survey}. The scalability problem can be handled by multiple components of air computing that can be reachable via seamless connection. Even though users are outside of the urban areas, air components can handle MAR tasks considering latency. Moreover, a huge amount of data can be transferred using the advantages of both 6G and 3D network structures.

\subsection{Sport and Concert Activities}
The communication infrastructure in terrestrial areas is built considering fixed resources in which once the facility is deployed, it can be changed with difficulty and significant additional cost. Considering capital expenditures (CAPEX) and operating expenses (OPEX) of companies, investing on fixed resources was plausible over years. For example, if there are limited resources such as base stations for users that increase in years, improving the capacity which means adding new base stations could be the solution for this problem. However, especially after the proliferation of the versatile application types in smartphones, the fixed infrastructure may not be sufficient for the user needs that change dynamically.

\begin{figure}[t]
\centering
\includegraphics[scale=0.13]{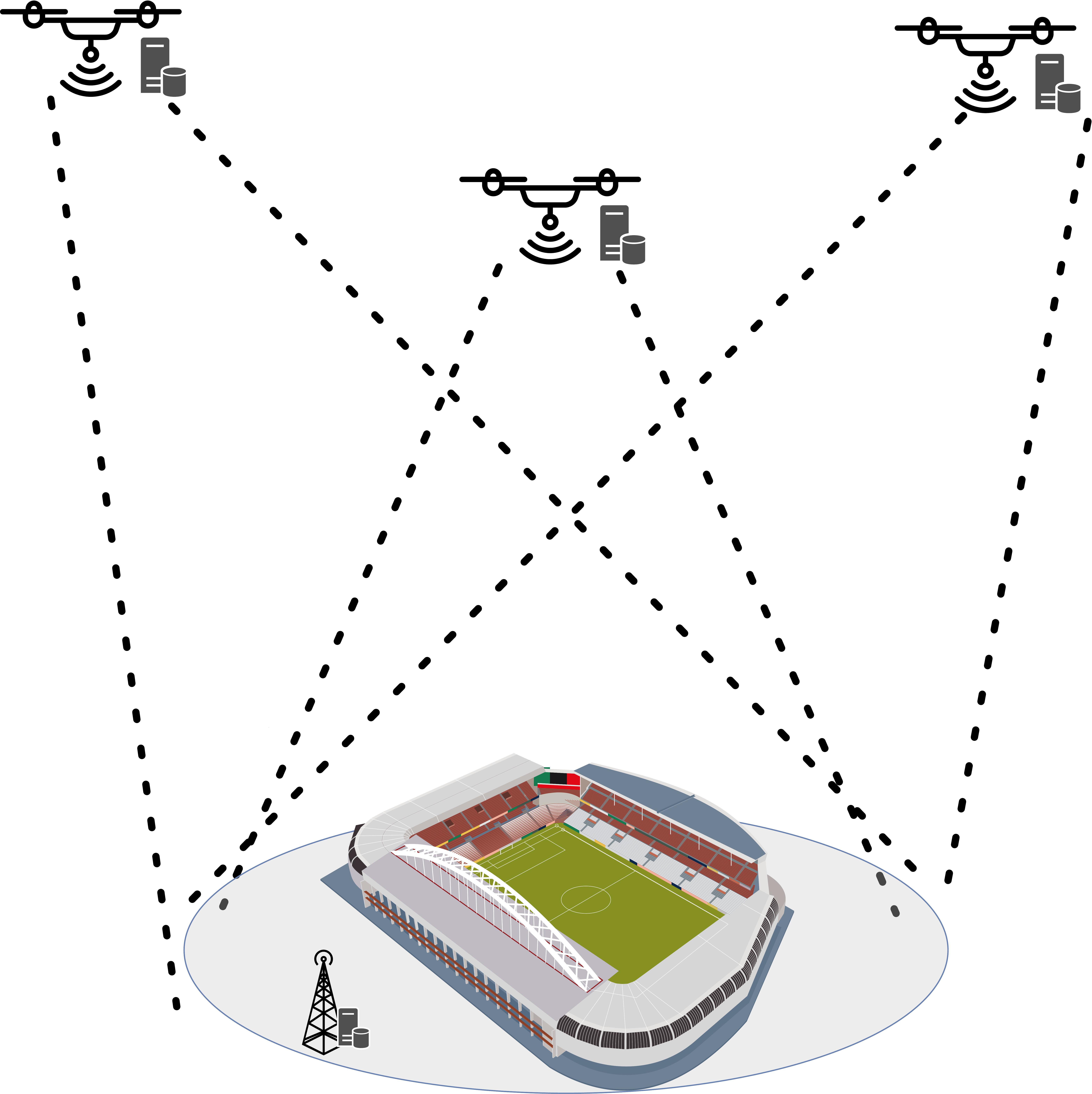}
\caption{Because of the crowded organizations such as sport events, the communication infrastructure would be insufficient to provide required QoS. Thus, air components can increase network capacity dynamically for such cases.}
\label{SportEvents}
\end{figure}

One of the most important examples of this situation is the sport and concert activities where thousands of people rally and use their applications. Statically built infrastructures expose a significant amount of traffic and computational offloading which may be an order of magnitude higher than the expected requests. Even though network slicing, network function virtualization (NFV), software-defined networks (SDN), and edge computing are used to provide a solution, the issue is still open. On the other hand, by using air components and directing them to the corresponding places where activities occur based on the current network requirements, we can increase the capacity of the network dynamically as shown in Figure \ref{SportEvents}. Hence, even though the network resources cannot meet the number of application requests, new resources and routes can be created through air computing. As a result, QoS and QoE would be enhanced.

\begin{figure*}[!t]
\centering
\includegraphics[scale=0.6]{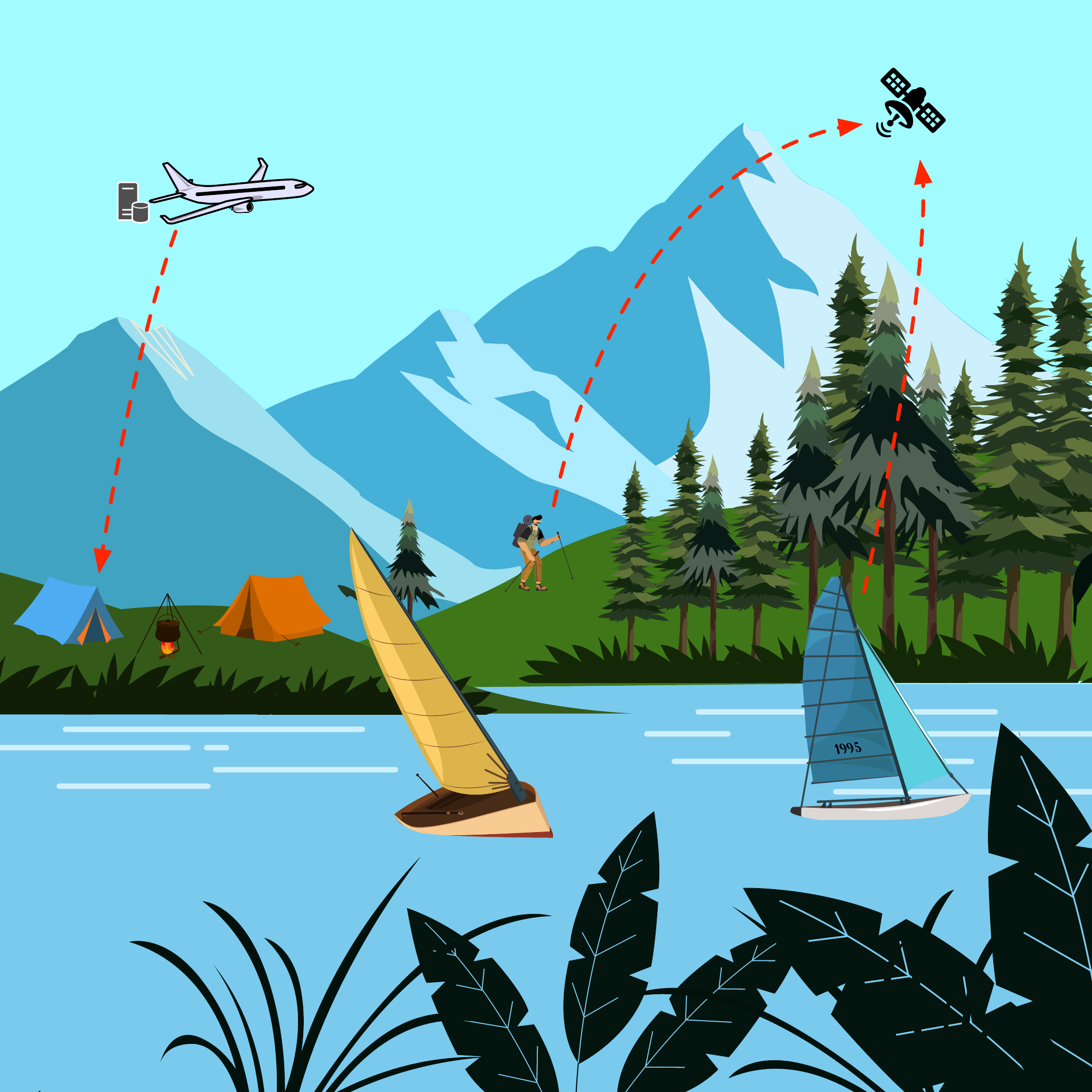}
\caption{Since there is a limited or no communication infrastructure in particular places for outdoor activities, the communication would continue through air computing.}
\label{Nature}
\end{figure*}
\subsection{Outdoor Activities}

Since people that perform outdoor activities including sailing, kayaking, climbing, and trekking may not access corresponding resources due to lack of communication infrastructure, they can face isolation in the natural environments. This situation would be crucial considering several issues. In the case of an injury or a health problem such as a heart attack, emergency services must be accessed immediately in order to obtain the aid. Regarding the current networking conditions, the communication from secluded areas including sea, mountain, and forest cannot be provided properly. Thus, even though there would not be such a problem, people carrying out the activities do not feel safe about those possible issues. Apart from the health, daily routines such as social media, communication via chat, and telephone call also cannot be carried out in those conditions. Hence, even though those activities provide important relief for people, their life quality may have deteriorated.

Through air computing, the issues for outdoor activities can be solved as shown in Figure \ref{Nature}. By deploying UAVs, HAP vehicles, and LEOs in suitable places, people in secluded areas can reach important contents and communication infrastructure. We assume that HAP and LEOs would be more useful for such activities as UAVs may face battery reload problems which is similar to the situation in rural areas.


\section{Components of Air Computing: Edge, LAP/HAP, LEO}
\label{Components-RelatedWork}

Air computing is made up of a variety of technological elements. In this section, we review individual elements, and discuss their benefits and contributions in the use-cases. Moreover, their open issues are also focused on and how air computing may serve as a remedy is put forward.


\subsection{Edge Computing}

Even though vertical networking adds another dimension to the current network infrastructure in air computing, the essential element in an urban area would be edge computing as it is deployed widely. As a result, determining a  profitable edge strategy is still important \cite{horner2021edge}. In this section, we investigate the most recent edge computing studies considering resource allocation, task offloading, edge caching, and energy efficiency issues. Moreover, we give an evaluation of selected edge studies in Table \ref{SelectedEdgeStudies} and analyze them considering the benefits of the air computing.

\subsubsection{Resource Allocation}
 Along with the computation/task offloading, resource allocation is one of the primary research issues in edge computing because of the new generation application requirements including transmission bandwidth, latency, energy consumption, and reliability \cite{horner2021edge, luo2021resource}. In \cite{dong2021joint}, authors perform intelligent task execution using DNN partitioning regarding heterogeneous edge server capacities. They propose a joint method considering cost-effective resource allocation and self-adaptive DNN partition in order to provide collaborative computation between IoT devices and edge servers. Lieang et al. focus on the challenge of handover between BSs considering the management of computation and radio resources \cite{liang2021multi}. Therefore, the goals in this study consist of maximizing the throughput and minimizing the handover cost. Chen et al. propose a cache-assisted multi-user MEC mechanism considering to cache executive codes of tasks proactively \cite{chen2021code}. Their goal was to reduce the task execution delay and energy consumption of users. Xia et al. consider the problems including resource allocation on demand for limited edge servers, and developing heterogeneous task offloading strategies \cite{xia2021online}. To this end, they implement an online distributed optimization algorithm based on game theory to perform optimal offloading and energy harvesting decisions. Bahreini et al. develop an auction-based mechanism by addressing the resource allocation and monetization challenges in MEC \cite{bahreini2021mechanisms}. They focus on the dynamic provisioning of computing resources since the tasks of users are heterogeneous. On the other hand, Roostaei et al. investigate Stackelberg game based distributed algorithm \cite{zhang2009stackelberg, maharjan2013dependable} in order to dynamically allocate and price edge resources \cite{roostaei2021game}. Zhao et al. develop a hybrid system considering beamforming and resource allocation \cite{zhao2020mobile}. They benefit from the advantages of mmWave communications indicated in \cite{wang2014cellular} in order to optimize beamforming vectors at users and base stations to minimize the maximum delay.


\begin{table*}[t]
\caption{Evaluation of Selected Edge Studies}
\label{SelectedEdgeStudies}
\centering

\begin{tabular}{ | p{0.5cm} | p{1.1cm} | p{3.5cm} |  p{2.9cm} | p{2.5cm} | p{4cm} |}
\hline
\textbf{Study} & \textbf{Category} & \textbf{Goal} & \textbf{Solution}  & \textbf{Open Issue} & \textbf{Benefits of Air Computing} \\
\hline
\cite{dong2021joint} & Allocation & Providing both computation efficiency and cost effectiveness to accelerate DNN-based task acceleration in the MEC & A joint method by a self-adaptive DNN partition with cost-effective resource allocation &  Only a single edge server is used & With multiple components, it can increase the capacity\\
\hline
\cite{liang2021multi} & Allocation & Maximizing the sum offloading rate, quantifying MEC throughput, and minimizing the migration cost & They relax the corresponding binary variables in the original problem to overcome non-convex issue &  Energy efficiency regarding edge servers is not considered & 3D network structure would alleviate the handover problem \\
\hline
\cite{chen2021code} &Caching & A cache-assisted multi-user MEC mechanism & They formulate a non-linear programming problem which involves a joint optimization &  There is no mobility & UAVs can be helpful for caching important contents \\
\hline 
\cite{yuan2021caching} &Caching & They investigate the cooperation problem of edge nodes in MEC & They use Lagrangian multipliers and then a distributed optimization algorithm &  There is no mobility and handover consideration & Vertical networking would increase capacity for caching based on air components. \\
\hline 
\cite{feng2021energy} &Offloading & A priority-differentiated offloading strategy that considers the stringent QoS requirements of mission-critical services & They use  Lyapunov optimization for priority-differentiation &  Only a single edge server is used & Mission-critical applications can be handled very well regarding multiple components of air computing. \\
\hline 
\cite{xue2021joint} &Offloading & Task offloading for multi-user and multi-vehicle in vehicular MEC & They propose a dynamic incentive mechanism &  Mobility model is not clear & Seamless connectivity can alleviate the problems in vehicular MEC systems \\
\hline 
\cite{bozorgchenani2020multi} &Energy & Minimizing both the energy consumption and task processing delay of the mobile devices & They propose an evolutionary algorithm that can efficiently find a representative sample of the best trade-offs &  There is no mobility and cloud consideration  & Multiple air components can alleviate the trade-offs between offloading and energy consumption \\
\hline 
\cite{chen2021energy} &Energy & Energy-efficient offloading for DNN based smart IoT systems & They propose a swarm optimization algorithm for energy-efficient offloading strategy&  There is no mobility & With the offloading to different nodes in the air, it may reduce energy consumption \\
\hline
\end{tabular}

\end{table*}

\subsubsection{Task Offloading}

Task offloading is the most crucial issue in edge computing regarding QoS of IoT devices and their corresponding applications \cite{laroui2021edge}. The performance of task offloading is generally evaluated with other metrics such as energy efficiency and edge caching hit \cite{dziyauddin2021computation}. Peng et al. used three constrained multiobjective algorithms considering time and energy consumption in order to solve the computation offloading problem in an edge environment \cite{peng2021constrained}. Feng et al. on the other hand focus on different requirements of mission-critical applications regarding their priorities \cite{feng2021energy}. To this end, they benefit Lyapunov optimization considering the energy consumption of resources \cite{neely2010stochastic}. However, they use only a single edge server in their experiments. Chen et al. on the other hand, develop a system that jointly optimizes task assignment and offloading scheduling in order to minimize maximum completion delay \cite{chen2021delay}. For this system, they also consider different communication and computation capabilities. Xue et al. develop a dynamic incentive mechanism to investigate the problem of the task offloading and resource allocation \cite{xue2021joint}. They consider a multi-user and multi-vehicle system. They use the Stackelberg game for the interaction between MEC service provider and UEs. In \cite{feng2021collaborative}, authors focus on data caching and computing offloading in a two-tier MEC environment. They consider the constraints of tasks in terms of the delay and the minimization of the network cost at user. However, they do not include mobility for the users and cloud option for offloading. Xu et al. investigate the performance of task offloading in high-speed railways (HSRs) considering proper data routing paths for each offloaded task \cite{xu2021throughput}. Since handovers are frequent in HSRs, they focus on how frequent handovers in uplinks and downlinks affect offloading. In \cite{zhang2021risk}, Zhang et al. focus on autonomous manufacturing by considering their delay sensitivity. To this end, they propose a risk-aware cloud-edge computing framework by developing a branch-and-check approach for solving the nonlinear programming problem. Yang et al. propose a machine learning solution to solve the offloading problem in MEC \cite{yang2020computation}. They train and jointly optimize the offloading decisions and resource allocation. Li et al. focus on caching techniques to optimize QoS in MEC \cite{li2021efficient}. They propose three algorithms to forecast the next executing task. Moreover, they jointly consider cache hit rate and load balance of edge servers.

\subsubsection{Energy Efficiency}



Since the battery capacity of IoT devices is limited, and service providers would like to lower their expenses regarding power consumption of edge servers, energy efficiency is an important research topic in edge computing. In \cite{bozorgchenani2020multi} authors focus on minimizing the energy consumption and task processing delay in this study. For that purpose, they develop an evolutionary algorithm that finds the best trade-offs between energy consumption and processing delay. Song et al. consider minimizing the energy consumption of mobile devices when executing corresponding tasks at satellites \cite{song2021energy}. Moreover, their model provides MEC services using LEOs for mobile devices in disaster areas. Chen et al. investigate energy-efficient offloading considering QoS requirements of DNN-based smart IoT systems \cite{chen2021energy}. To this end, they design a self-adaptive particle swarm optimization algorithm for the corresponding energy-efficient offloading strategy. In \cite{ghoorchian2020multi}, authors develop a multi-armed bandit algorithm to provide a solution for the server selection problem in edge computing. They define corresponding reward and cost terms considering the energy and required time in offloading rounds. Zhou et al. focus on energy-efficient service migration in considering MEC-enabled dense cellular networks \cite{zhou2021energy}. They formulate the service migration process as a MINP problem. They also use the Lyapunov optimization technique to decouple the migration process.

\subsection{LAP}





Since providing Line of Sight (LoS) links has many advantages in terms of connectivity, service provision, and latency, UAVs have been deployed in many areas especially remote locations. However, this deployment also brings its own issues including mobility management, UAV networking management, and flight formation \cite{lyu2018uav, gupta2015survey}. To this end, we categorize and evaluate these issues under trajectory planning, task offloading, placement of UAVs, and energy consumption.

\subsubsection{Trajectory Planning}




In order to provide efficient on-demand services, trajectory planning is crucial for UAVs \cite{shi2019multi, zhou2015multi}. Moreover, optimization of the pre-defined paths based on the dynamic events is also critical for the performance of UAVs in terms of QoS \cite{jeong2017mobile}. Zhao et al. investigate a proactive mobility management solution for users' trajectories in order to deploy UAVs dynamically in the network \cite{zhao2021predictive}. To this end, they propose a distributed learning framework in which edge servers are considered as local data owners that collect connection data. Wang et al. aim at minimizing the energy consumption of users in the network by considering the resource allocation and trajectory of UAVs \cite{wang2021deep}. They propose two solutions: (1) convex optimization based trajectory control algorithm to minimize energy consumption, and (2) DRL-based trajectory control algorithm for real-time decisions. Similarly, authors in \cite{li2019joint} propose that the trajectory of UAVs can be approximated using traditional convex optimization approaches and discrete variables. Liu et al. optimize UAV trajectories considering energy consumption \cite{liu2020path}. They formulated the problem as MDP and proposed DRL with a double q-network.  Wang et al. proposed a multi-UAV communication system for 6G in which they consider UAV trajectories and radio resource scheduling \cite{wang2020collaborative}.

\subsubsection{Task Offloading}

One of the most important motivations for the deployment of UAVs is the computation rate maximization of applications by using task offloading \cite{zhou2018computation, motlagh2016low}. Through task offloading via Line of Sight, the burden on the edge servers would be alleviated and required QoS can be provided. Seid et al. study on minimization of the computation costs in terms of energy consumption and computation delay \cite{seid2021multi}. They propose a multi-agent deep reinforcement learning (MADRL)-based approach in a multi-UAV enabled IoT edge network using a single centralized SDN controller. In \cite{apostolopoulos2021data}, the authors propose an offloading system in which users can perform partial offloading regarding UAVs and MEC servers. They formulate the problem as a maximization problem and use the principles of Prospect Theory \cite{kahneman2013prospect}. Haber et al. on the other hand focus on mission-critical applications that require ultra-reliable low-latency computation offloading \cite{el2021uav}. They use UAVs considering the maximization of served request rate and the optimization of UAVs' positions with the offloading decision. Zhan et al. propose a framework for a multi-UAV enabled MEC system in order to maximize the number of served IoT devices regarding computation offloading and resource allocation \cite{zhan2021multi}. Zhao et al. proposed a collaborative task offloading approach in a multi-UAV multi-MEC system considering energy consumption, and UAV trajectory \cite{zhao2022multi}. For this purpose, they use a cooperative multi-agent DRL (MADRL) method in which the policy gradient algorithm is utilized. Diao et al. investigate the usage of UAVs as relay nodes considering emergency conditions \cite{diao2021uav}. Moreover, they consider energy consumption minimization by optimizing offloading and scheduling.

\subsubsection{Energy Consumption}

Even though energy consumption is considered as a performance metric that is evaluated in studies along with the other issues such as task offloading, and trajectory planning, some studies take energy efficiency into account as the main problem. Ji et al. consider nonorthogonal and orthogonal multiple access modes for a UAV-assisted MEC system and focus on weighted-sum energy consumption \cite{ji2020energy}. To this end, they propose alternating iterative algorithms in order to optimize UAV trajectory and resource allocation. The goal of Li et al. is to create a model for UAV-assisted MEC by considering energy-efficient UAV trajectory design and optimized computation offloading \cite{li2020energy}. They also consider partial offloading in this study. Chen et al. focus on ultra-dense networks considering the resource allocation problem by maximizing energy efficiency \cite{chen2021deep}. They used UAVs as flying base stations (BS) and utilized DQN as the solution technique. Liu et al. aim to minimize the energy consumption of users by optimizing relay and computing features of UAVs \cite{liu2021energy}. They use an iterative algorithm to solve the non-convex problem.

\subsubsection{Placement}




The placement of UAVs is substantial as coverage provides connectivity that enhances the network capacity in terms of the data transmission and computation \cite{borralho2021survey}. Moreover, their utilization in poorly covered terrestrial regions would increase end-users QoE \cite{mozaffari2016unmanned}. Therefore, placement optimization would be crucial for the performance of the UAV-based network along with the trajectory planning \cite{lyu2016placement, alzenad20173}. 

In \cite{liu2018energy}, Lui et al. use actor-critic methods in DRL in order to provide connectivity between UAVs so that they can cover required areas to improve QoS. Wang et al. optimize the placement of the UAVs considering the offloading decision and resource allocation in a multi-UAV-enabled MEC environment \cite{wang2019joint}. They propose a two-layer optimization method to solve the problem. On the other hand, Yuan et al. focus on the dynamic placement of UAVs in a vehicular network \cite{yuan2021harnessing}. They utilize the actor-critic DRL approach to carry out real-time UAV placement. They also consider UAVs' flying range, communicating range, and energy resources.

\begin{table*}[t]
\caption{The Summary of Main Differences Between Air Layers}
\label{DifferencesBetweenAirLayers}
\centering

\begin{tabular}{ | p{2.5 cm} | p{4.5 cm} | p{4.5 cm} |  p{4.5 cm} | p{4.5 cm} |}
\hline
\textbf{Issue} & \textbf{LAP} & \textbf{HAP} & \textbf{LEO}\\
\hline
Altitude & Less than 10km & 10 - 30km & 160 - 2000km\\
\hline
Propagation Delay & 10 - 30 $\mu$s & 50 - 85 $\mu$s & 1.5 - 3ms\\
\hline
Coverage Scale & Cell & Regional & Continental\\
\hline
Main Deployment & Urban Areas & Urban and Suburban Areas & Rural Areas\\
\hline
Low Latency Apps & It can be provided since propagation delay and 6G wireless communication can ensure the corresponding requirements. & It depends on the channel and weather conditions regarding propagation delay.  & Because of the high altitude and propagation delay, satellites cannot provide the requirements of low latency applications.\\
\hline
Main Use-cases & Seamless mobility is ensured through UAVs. Moreover, components in this layer provide either edge computing solutions or access to edge servers. & Airplanes and balloons can be used as management nodes for UAVs considering their regional coverage. Furthermore, they can also be used for edge computing purposes. & Satellites can perform edge computing solutions however their service would be limited due to their low on-board capacity. Therefore, they are generally used to access the cloud computing solutions.  \\
\hline
Performance & As UAVs can be configurable easily regarding their stationary position, they can use their capacity effectively based on the user density. & Even though their configurability is not flexible as UAVs, balloons and aircrafts may provide a relative stationary position. However, they cannot use their capacity as efficient as UAVs. & Due to their high speeds and their deployment in underpopulated areas, some of the capacity of satellites would be wasted. \\
\hline
Maintenance & Even though UAVs provide important flexibility in dynamic environments, they need charging stations. Moreover, their maintenance would be daily due to their limited battery capacity. They can be reusable multiple times. & Balloons and aircrafts can fly for days based on their fuel capacity. However, they must return to their corresponding bases for maintenance. They can be reusable multiple times. & Satellites are not recoverable after their deployment. However, they can give service for months. \\
\hline
Energy Consumption & The required energy is ensured from batteries. Their energy consumption can be heavily affected by winds and weather conditions especially if they fly against the wind. & The required energy is provided from fuels. The effect of the winds and weather conditions to the energy consumption is limited due to their altitude. & They meet the required energy from the solar power and corresponding batteries. Their energy consumption cannot be affected by weather conditions.\\
\hline
\end{tabular}

\end{table*}

\subsection{HAP Components and LEO Satellites}


Considering the limited battery energy and cell-based coverage of UAVs, HAP and LEO layers provide important advantages regarding long-distance communication, energy consumption, and management opportunities. Moreover, their performance would not be affected by the weather conditions due to their high altitudes as 10 - 30km for HAP components, and 160 - 200km for LEO satellites.

The main goal of the deployment of the HAP components is to provide connectivity such as Internet access, and to perform as a controller node for the UAVs  \cite{kurt2021vision, qiu2019air}. However, in certain circumstances, such as a congestion in the lower layers, they can be used as an edge computing server or a relay node. Since their coverage is on a regional scale, UAVs that can be considered  as dynamic cells can reach the corresponding resources in a particular area via HAP components. Even though this is one of the reasons that the deployment of HAP components is generally in urban and suburban areas, maintenance issues based on  energy consumption also cause an important restriction for their deployment in rural areas.

\begin{figure}[t]
\centering
\includegraphics[scale=0.042]{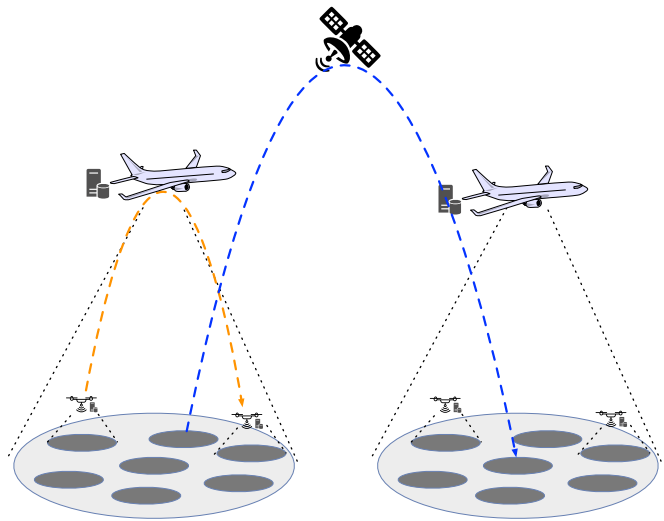}
\caption{While HAPs can provide regional access and management to LAP components, LEO satellites may ensure inter-regional and continental access for the corresponding resources.}
\label{Sagin}
\end{figure}

On the other hand, the essential use case of the LEO satellites is to meet the coverage problems of rural areas where infrastructure for the communication is insufficient. They can be deployed for months thanks to their efficient energy consumption, however they are not recoverable after their deployment. Moreover, they cannot be used for low latency applications due to their high propagation delay. Therefore, they can be used as a complementary resource for urban and suburban areas in order to get access to cloud computing solutions in WAN. Besides, exploiting services in continental or beyond regional distances would be more efficient using LEO satellites since terrestrial nodes may cause high latency \cite{kodheli2020satellite}. Figure \ref{Sagin} depicts reaching regional and inter-regional resources using HAP and LEO layers. 

It is important to note that when the LEO layer is exploited, SAGIN is the term that is generally used by studies to indicate the corresponding communication system \cite{liu2018space}. In \cite{tang2021deep}, Tang et al. developed an efficient offloading mechanism for SAGIN. They benefited from the communication between the LEO satellite network, LAP vehicles, and terrestrial resources. To minimize the total delay and to handle the high mobility of nodes, they proposed a deep reinforcement learning traffic offloading approach. Chen et al. \cite{chen2022satellite} benefit from satellite constellations to apply Federated Learning considering communication overhead and privacy issues. They use four different modes including remote cloud learning, onboard satellite learning, federated learning with data sharing, and federated learning with no data sharing to evaluate the performance of their system. On the other hand, Guo et al. focus on service coordination between different air layers in SAGIN \cite{guo2021service}. Therefore they separate the requirements of the environment using three service coordination scenarios: (1) fine-grained, (2) medium-grained, and (3) coarse-grained. In the fine-grained scenario, the network in the air is used as complementary regarding the need for the terrestrial network and ubiquitous coverage. Considering the delay-sensitive applications, coordination of the data processing and data communication services is provided by using medium-grained coordination. Finally, mobility and the corresponding service migration are ensured using coarse-grained service coordination.

In \cite{zhou2019delay}, Zhou et al. investigate dynamic scheduling problem in task offloading considering SAGIN environment. They deploy UAVs as flying gateways in order to perform the offloading decision. Considering the dynamic environment, they formulate the corresponding problem as an MDP and then apply linear programming. Similarly, Cheng et al. focus on computational offloading in SAGIN considering energy and computation constraints \cite{cheng2019space}. In their design, UAVs provide edge computing while satellites ensure access to cloud computing. To learn the optimal policy in a dynamic environment regarding large action space and mobility, they use an actor-critic DRL algorithm. In \cite{zhang2019satellite}, Zhang et al. analyze the architecture and corresponding application scenarios of satellite MEC. They propose network function virtualization and cooperative task offloading methods in order to integrate computing resources and improve the efficiency regarding delay and energy consumption.

Resource allocation and controller placement problems are also essential for LEO studies. In \cite{chen2021mobility}, Chen et al. examine the dynamic assignment and placement of controllers in SDN-based LEO satellite networks. They consider two challenges including highly dynamic topology, and randomly fluctuating load. To this end, they take propagation and queueing delays into account and then formulate the adaptive controller placement and assignment problem considering management costs. On the other hand, Zhang et al. investigate resource allocation in non-orthogonal multiple access (NOMA) terrestrial-satellite networks in which terrestrial and satellite components use the same spectrum for the communication \cite{zhang2022resource}. Since the original optimization problem is non-convex, they divided the original problem into three subproblems including user association, bandwidth assignment, and power allocation.

Since each layer is crucial for the performance of an air computing environment, we also summarize their essential features in Table \ref{DifferencesBetweenAirLayers} based on the critical issues. Thus, which layer should be used for particular requirements can be more distinct.

\section{Challenges, Opportunities and Future Research Directions}


Even though air computing can solve many issues related to the limitation of current networking paradigms, it would face many challenges such as network architecture, regulation of air vehicles, battery issues, coverage, and a communication protocol. Considering the fact that UAV communication between different devices causes many challenges \cite{fotouhi2019survey}, applying different vehicles in the air and providing communication between them is not trivial. 

Therefore, all of these issues must be investigated thoroughly in order to apply the air computing paradigm correctly. On the other hand, these challenges open new research areas regarding the provision of QoS, energy efficiency, determining air vehicle placement, and the deployment of AI. In this section, we elaborate on those challenges and corresponding research opportunities.

\subsection{Air Computing Architecture Design}


Since air computing has a 3D structure with four major layers including terrestrial, LAP, HAP, and LEO, the networking architecture in terms of offloading mechanism and routing is crucial \cite{huang2019airplane}. One of the main concerns in networking is how the requests would be handled considering the mesh connected  different nodes in the 3D structure. To investigate this, we evaluate three candidate design approaches including hierarchical, free, and hybrid designs. Moreover, we also make a comparison between a distributed approach and orchestration in air computing regarding task offloading.


\subsubsection{Hierarchical Design}

\begin{figure}[t]
\centering
\includegraphics[scale=0.09]{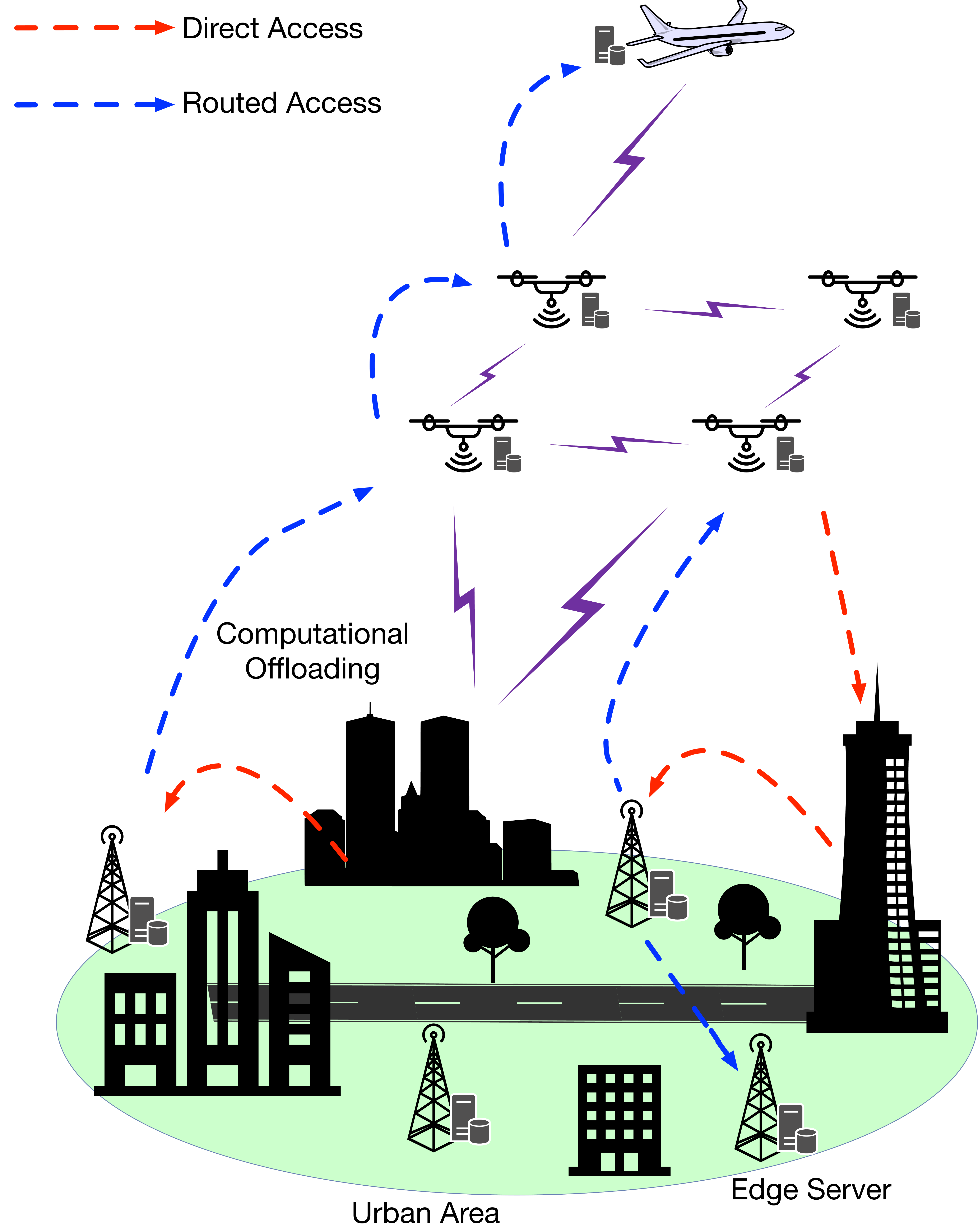}
\caption{Architecture of the hierarchical design.}
\label{HierarchicalDesign}
\end{figure}

As the name suggests, there is a systematic order between air computing layers in the hierarchical design as shown in Figure \ref{HierarchicalDesign}. If a task is offloaded from a user in a terrestrial network, the selection of the corresponding server for the computation is handled by another entity in the air computing environment, not by the user. For example, if the service for the corresponding application task can be met in one of the HAP vehicles due to network conditions, the task is routed regarding a predefined path rather than directly transmitted to the corresponding HAP component by the user. Therefore, the task is first sent to the nearest edge server in LAN. If the edge server cannot process the task because of its high load or its service incapability, the task is relayed to one of the UAVs in LAP. Note that we assume that the corresponding edge server is in the vicinity of the UAV. Next, if the UAV similarly cannot execute the task, it relays the task to one of the UAVs in its vicinity, one of the available edge servers on the ground, or one of the airplanes in HAP. Since the corresponding service has been given in HAP in this case, the task is pushed to the HAP. 

Even though the hierarchical design provides important manageability throughout the air computing environment, it may face high delays due to its layered structure. Based on the use cases and user profiles, it can be applied in the network for specific areas.

\subsubsection{Direct Access Design}

\begin{figure}[t]
\centering
\includegraphics[scale=0.09]{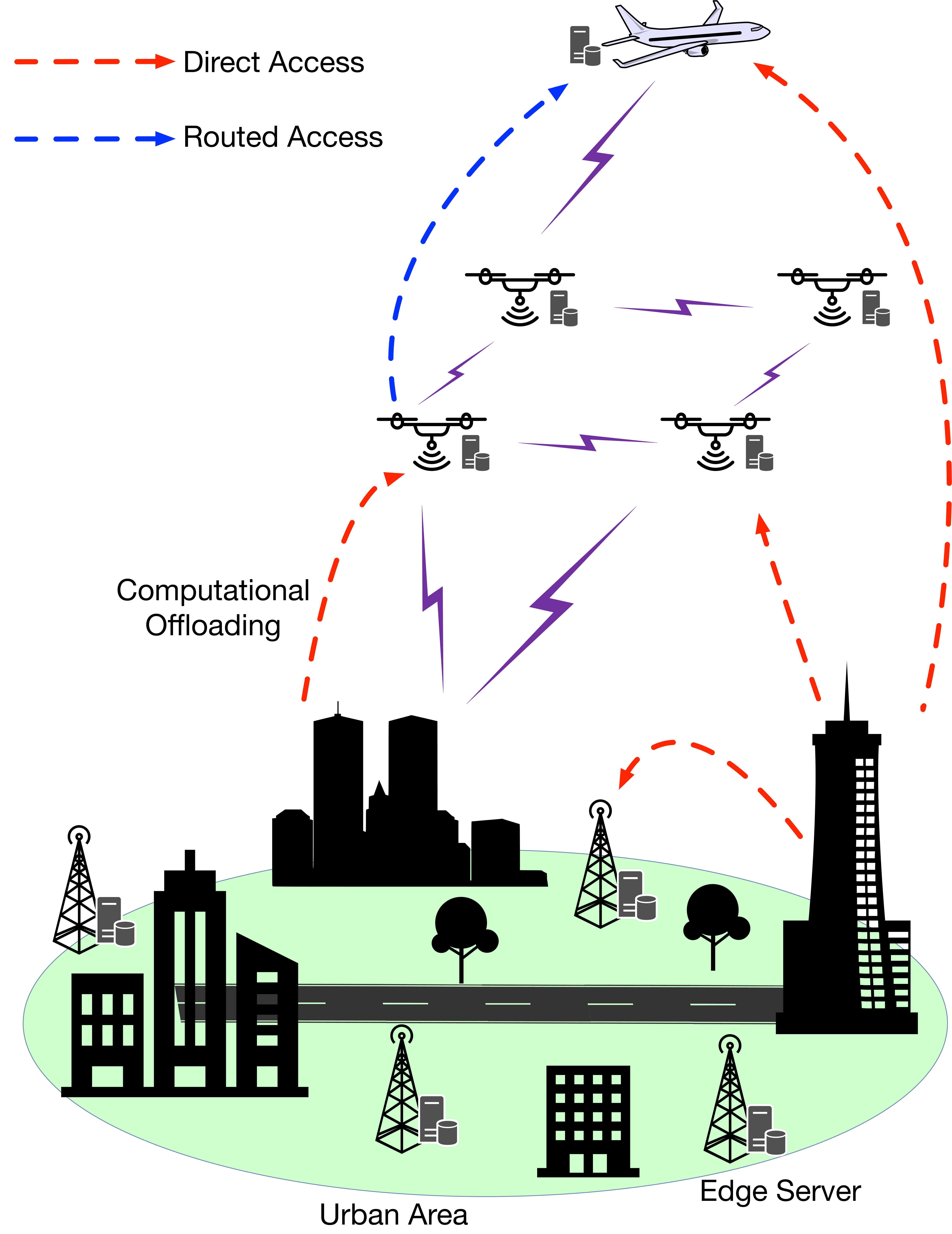}
\caption{Architecture of the direct access design.}
\label{FreeDesign}
\end{figure}

In contrast to the hierarchical design, in direct access design, IoT devices in an air computing environment can select the corresponding server to meet the requirements of their tasks as shown in Figure \ref{FreeDesign}. As a result, if the corresponding sender knows which server provides the required service, it offloads the task directly to the appropriate layer through the seamless connectivity feature of air computing. 

However, this free access for each device in the network would cause congestion and underutilization of the resources. In terms of congestion, if a service is given by a particular server in the network, all devices which require the service may offload their task to that server. As a result, communication links and server capacity would be heavily affected by this situation. On the other hand, if devices use particular servers in the network due to low delay, high processing, and energy efficiency, some resources would be underutilized.

\subsubsection{Hybrid Design}

Considering the advantages of hierarchical design and direct access design, a hybrid design would be applied in an air computing environment. For the urban areas, in which the network is extremely dense regarding both end-devices and corresponding servers, the application of hierarchical design would be more suitable in order to prevent  underutilization and congestion cases. Moreover, the delay issue in hierarchical design can be covered by diverse resources regarding the provision of different services, and seamless connectivity through 6G. 

On the other hand, considering the suburban and rural areas, the utilization of direct access design would be more convenient since the infrastructure in those areas is limited. Therefore, direct access to the corresponding servers can be of benefit in terms of delay.

\subsection{Request Management}

In an air computing environment, handling user and IoT requests are crucial to meet the required QoS. Moreover, since there are different layers in the air, the request management would be more complex than other networking paradigms such as edge computing. Even though architecture design is essential for request management, deciding where to offload and when to offload is crucial for the performance. To this end, the benefits of a distributed system and an orchestrator-based system must be investigated thoroughly.

A distributed system in air computing can be described such that the device which offloads the task takes the decision of where to offload. For example, users can select one of the edge servers for offloading, or an edge server relays the offloaded task to one of the UAVs without consulting any intermediate device. One of the most important advantages of a distributed system is its scalability and its low latency. Since there is no intermediate element considering the offloading, the transmission of the request would be faster.

On the other hand, it has a significant drawback considering the current state of the network. An offloading device in the environment cannot be aware of the current condition of the corresponding servers and network elements in terms of the load of the servers, and the number of requests that may affect the communication links. Therefore, if there is no additional communication regarding this information, the decision taken by the offloading device would cause a failed task offloading. Moreover, note that if an additional communication mechanism is deployed in the environment, it must be well-optimized so that the communication links cannot be affected by the transmission regarding congestion.

In contrast to distributed systems, an orchestrator to which the tasks are first sent by the offloading devices can be used in different parts of an air computing environment. In such a system, a user can directly offload its task to the orchestrator, and the decision of where to offload can be taken by the orchestrator itself which is aware of the current condition in the network regarding communication links and corresponding servers.

Even though it has advantages, several points must be investigated and optimized in an orchestrator-based system. First, the deployment of the orchestrator is crucial for the performance of the system since many entities in the network may send their corresponding requests. Therefore, it should be reachable in a suitable time so that it can relay those requests without violating the maximum delay requirement based on application type. Second, the centralized deployment of an orchestrator results in a single point of failure such that the corresponding portion of the network would be heavily affected as users and IoT devices cannot offload their tasks. To this end, a fault management system should be considered with the deployment of an orchestrator. Finally, third, the cost of an orchestrator in terms of new communication links, delay, and congestion must be minimized for the system performance. 



\subsection{Deploying Artificial Intelligence}

Considering the diverse requirements of various applications and IoT devices in an air computing environment, traditional optimization-based solutions and heuristics would provide limited solutions. Therefore, the application of AI-based solutions will be inevitable. Especially, regarding edge and UAV-based systems, there are already many studies which benefit from AI-based solutions including ML, DL, and DRL \cite{wang2020convergence, chen2020deep, sun2021cooperative}. As a result, along with the requirements of the air computing paradigm, novel AI solutions should be applied to meet new challenges.

Since there are many resources that produce an enormous amount of data in an air computing environment, processing them and then learning meaningful patterns considering the performance of the system can be feasible using ML techniques. However, in recent years, DL solutions have been preferred rather than traditional ML algorithms since DL is more successful in terms of training, non-linear transformation, efficiency, and required computing power \cite{lecun2015deep}.

Even though DL solutions are preferred in recent studies, the data collection phase in an air computing environment would cause serious degradation of system performance. First, the huge amount of data can bring about congestion problems on communication links. Second, for such a heterogeneous environment, providing the privacy of the data would be difficult. Therefore, applying Federated Learning (FL) solutions would be more suitable with air computing \cite{nguyen2021federated, samarakoon2018federated}. 

On the other hand, since many decisions must be taken based on dynamic events in the air computing environment, and they may not be labeled due to the nature of the problem, supervised ML and DL based solutions would be inadequate. To this end, recent studies benefit from DRL in which the agent can learn directly from the environment without needing human interaction \cite{chen2021deepsurvey}. The agent in an air computing environment can be the orchestrator, UAV, or edge server since they take actions based on the current state of the system. Thus, we believe that future studies can consider the deployment of DRL solutions along with FL.

\subsection{Air Computing Protocol} 

As there are many different entities in the air computing environment, the communication between them should be carried out based on predefined rules, which are defined by a protocol. An air computing protocol should be reliable, secure, and fast so that the entities carry out communication easily \cite{hassija2021fast}. Moreover, it should facilitate the management of the network as it must provide data integrity.


\subsection{Energy Issue of Air Vehicles}

The vehicles in air layers use either batteries or fuel. Therefore, their deployment, trajectories, and computing capacities should be well-optimized to ensure energy efficiency. As mentioned in Section \ref{Components-RelatedWork}, energy-related issues are currently evaluated regarding edge and UAV studies. However, considering all layers of air computing, a collaboration between different air vehicles can reduce energy consumption further.

\subsection{Flying Vehicle Regulations} 


Since each country has different regulation policies considering flying air vehicles, the entities in an air computing environment should comply with those corresponding rules. Moreover, the air computing protocol should also be in compliance with regulations as reliable and fast communication is vital for flying entities in the air. On the other hand, considering the existing cellular network infrastructure, the standardization between existing resources and newly deployed air computing devices must be investigated for 6G and beyond \cite{abdalla2022open, bor20195g}.

\subsection{Movement and Coverage of Air Vehicles}

Even though deployment is an important research issue for 2D terrestrial networks, this problem is more difficult to manage as air vehicles move. Moreover, since their coverage, transmission quality, and power consumption are heavily affected by their vertical movement, optimization of their altitude and trajectory is crucial \cite{he2017joint, huang2018joint}. Therefore, request management can be handled along with this optimization in order to provide efficient performance.

\section{Conclusion}

In this study, we defined a novel, next-generation computational paradigm called air computing. In the face of ever-growing application resource demands, air computing strives to solve bottlenecks and inefficiencies of the computational infrastructure by intelligently harmonizing 2D legacy terrestrial resources with novel 3D vertical networking technologies. 

Air computing is indeed based on a family of technologies such as UAV, LAP, HAP, LEO, 6G, and edge computing. In order to give a complete overview, we first investigated air computing as a whole regarding its main features and how it contrasts with former systems such as edge and cloud computing. We then described the individual technological components and how they fit in the overall architecture. A detailed literature review for the individual components is also provided to give a full technical overview of air computing in all aspects.

Moreover, we examined the advantages that would be put forward by a possible air computing implementation regarding the QoS requirements of the next-generation applications and QoE expectations of end-users. Then, we elaborated on the potential use cases where the current paradigms experience difficulty in meeting the dynamic user demands. Finally, we analyzed the opportunities and the corresponding challenges in an overall context from the perspectives of both the end users and service providers. Inspired by the challenges involved, we presented a selection of future research directions which we believe have the strong potential to transform the domain.

\section*{Acknowledgment}
This work is supported by the Turkish Directorate of Strategy and Budget under the TAM Project number 2007K12-873.

\bibliographystyle{IEEEtran}
\bibliography{AirComputing}

%

\begin{IEEEbiography}
[{\includegraphics[width=1in,height=1.25in,clip,keepaspectratio]{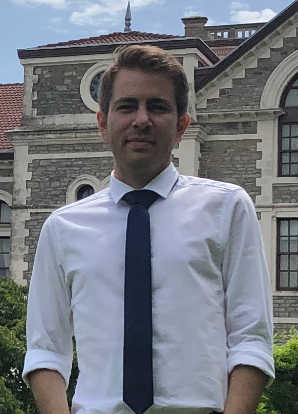}}]{Baris Yamansavacilar}
received his BS degree in Computer Engineering from Yildiz Technical University, Istanbul, in 2015. He received his MS degree in Computer Engineering from Bogazici University, Istanbul, in 2019. Currently, he is a PhD candidate and a research assistant in Computer Engineering Department at Bogazici University. His research interests include Edge Computing, Deep Reinforcement Learning, Mobile Networks, Software-Defined Networking, and Machine Learning.

\end{IEEEbiography}

\begin{IEEEbiography}[{\includegraphics[width=1in,height=1.25in,clip,keepaspectratio]{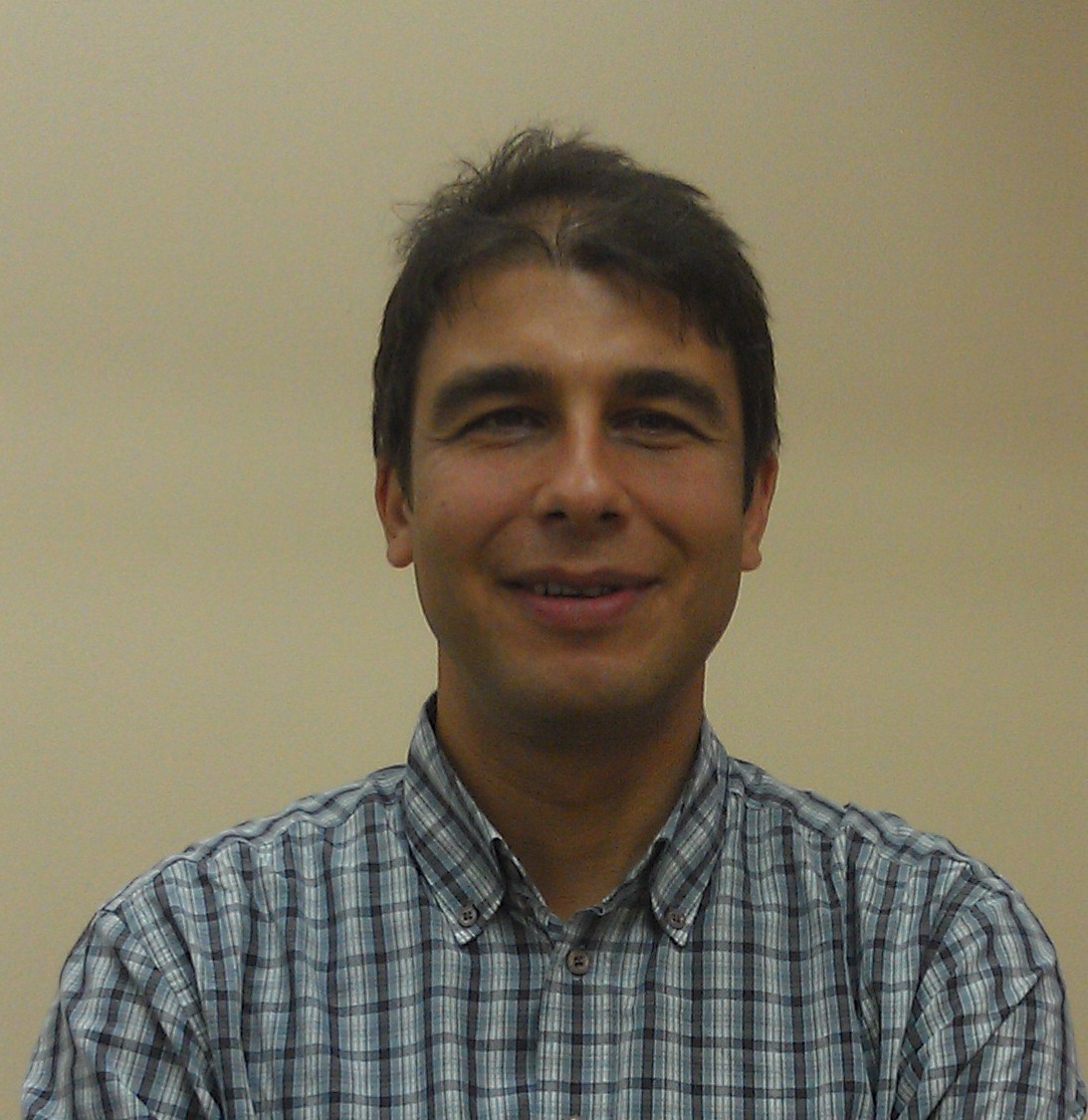}}]{Atay Ozgovde}
received the BS and MS degrees from Bogazici University, Istanbul, in 1995 and 1998, respectively. In 2002, he started working as a research assistant in the Computer Engineering Department, Bogazici University, where he completed the PhD degree in the NETLAB research group in 2009. He worked as a researcher at the TETAM research center to complete his postdoctoral research where he worked with the WiSE-Ambient Intelligence Group. Currently, he is an assistant professor in the Computer Engineering Department, Bogazici University. His research interests include wireless sensor networks, embedded systems, distributed systems, pervasive computing, SDN and mobile cloud computing. He is a member of the IEEE.
\end{IEEEbiography}

\begin{IEEEbiography}[{\includegraphics[width=1in,height=1.25in,clip,keepaspectratio]{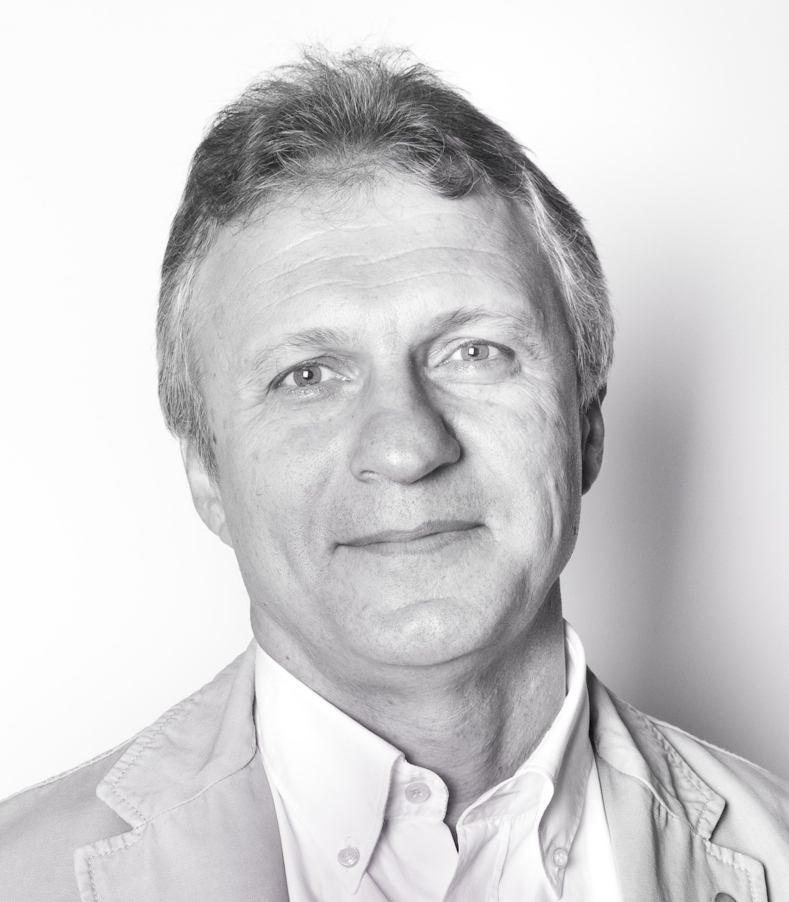}}]{Cem Ersoy}	
worked as an R\&D engineer in NETAS A.S. between 1984 and 1986. After receiving his PhD from Polytechnic University, New York in 1992, he became a professor and currently the department head of Computer Engineering in Bogazici University. Prof. Ersoy's research interests include wireless/cellular/adhoc/sensor networks, activity recognition and ambient intelligence for pervasive health applications, green 5G and beyond networks, mobile cloud/edge/fog computing, software defined networking, infrastructureless communications for disaster management. Prof. Ersoy is also the Vice Director of the Telecommunications and Informatics technologies Research Center, TETAM. Prof. Ersoy is a member of IFIP and was the chairman of the IEEE Communications Society Turkish Chapter for eight years.
\end{IEEEbiography}

\end{document}